%% file: main.tex
\documentclass{article}

\usepackage{microtype}
\usepackage{graphicx}
\usepackage{subfigure}
\usepackage{booktabs} 
\usepackage{amsmath,amssymb,amsfonts} 
\usepackage{multirow} 
\usepackage{adjustbox} 
\usepackage{hyperref}
\usepackage{xcolor}

\usepackage[accepted]{mlsys2025}
\mlsystitlerunning{MAS-Attention: Memory-Aware Stream Processing for Attention Acceleration on Resource-Constrained Edge Devices}

\begin{document}
\twocolumn[
\mlsystitle{MAS-Attention: {M}emory-{A}ware {S}tream Processing for Attention Acceleration on Resource-Constrained Edge Devices}



\begin{mlsysauthorlist}
\mlsysauthor{Mohammadali Shakerdargah}{ualberta,huawei}
\mlsysauthor{Shan Lu}{huawei}
\mlsysauthor{Chao Gao}{huawei}
\mlsysauthor{Di Niu}{ualberta}
\end{mlsysauthorlist}

\mlsysaffiliation{ualberta}{Department of Electrical and Computer Engineering, University of Alberta, Edmonton, Canada}
\mlsysaffiliation{huawei}{Huawei Technologies, Edmonton, Canada}

\mlsyscorrespondingauthor{Mohammadali Shakerdargah}{shakerda@ualberta.ca}
\mlsyscorrespondingauthor{Di Niu}{dniu@ualberta.ca}

\mlsyskeywords{Machine Learning, MLSys}

\vskip 0.3in

\begin{abstract}
\input{sections/abstract}
\end{abstract}
]



\printAffiliationsAndNotice{}  

\section{Introduction}
\input{sections/introduction}
\section{Related Work}
\input{sections/background}

\section{MAS-Attention Overview}
\input{sections/challenge_motivation}

\section{Methodology}

\input{sections/methodology}

\section{Experiments}
\input{sections/experiment_2}

\section{Conclusion \& Future Work}
\input{sections/conclusion}

\nocite{langley00}

\bibliography{refs}
\bibliographystyle{mlsys2025}

\end{document}

%% file: sections/abstract.tex
\label{sec:abstract}

The advent of foundation models have revolutionized various fields, enabling unprecedented task accuracy and flexibility in computational linguistics, computer vision and other domains. 
Attention mechanism has become an essential component of foundation models, due to their superb capability of capturing correlations in a sequence.
However, attention results in quadratic complexity in memory and compute as the context length grows.
Although many fusion-based exact attention acceleration algorithms have been developed for datacenter-grade GPUs and accelerators leveraging multi-core parallelism and data locality, yet it remains a significant challenge to accelerate attention on resource-constrained edge neural accelerators with limited compute units and stringent on-chip caches. 
In this paper, we propose a scheme for exact attention inference acceleration on memory-constrained edge accelerators, by parallelizing the utilization of heterogeneous compute units, i.e., vector processing units and matrix processing units.
Our method involves scheduling workloads onto these different compute units in a multi-tiered tiling scheme to process tiled vector workloads and matrix workloads in attention as two streams, respecting the workload dependencies. We search for tiling factors to maximize the parallelization of both compute units while considering I/O overhead, and propose a proactive cache overwrite strategy to avoid undesirable cache spills in reality. 
Extensive results based on open-sourced simulation frameworks show up to 2.75$\times$ speedup and 54\% reduction in energy consumption as compared to the state-of-the-art attention fusion method (FLAT) in the edge computing scenario.   
Further experiments on a real-world edge neural processing unit demonstrate speedup of up to 1.76$\times$ for attention as compared to FLAT, without affecting model output accuracy.

%% file: sections/introduction.tex
\label{sec:introduction}

   Foundation models \cite{vaswani2017attention, kitaev2020reformer, kaplan2020scaling, peebles2023scalable, li2024multimodal} have driven recent advancements in generative AI on edge devices such as smartphones, especially in AI agents \cite{zhang2023appagent, wang2024mobile, fan2025videoagent}, large language models (LLMs) \cite{radford2018improving, ouyang2022training, glaese2022improving, mehta2024openelm} and text-to-image diffusion models \cite{poole2022dreamfusion, esser2024scaling}. 
    Central to these models is the attention mechanism, which captures long-range dependencies between tokens, but incurs quadratic memory and computational complexity due to pairwise token interactions. Deploying these models is challenging, especially on resource-constrained edge devices with limited on-chip cache and processing power. 

    Significant efforts have been made to accelerate attention computation through software fusion techniques on datacenter-grade hardware. For cloud servers, multi-core parallelism \cite{shoeybi2019megatron, rasley2020deepspeed, narayanan2021efficient, kwon2023efficient, liu2023ring, cho2024kv} and efficient utilization of on-chip SRAM in GPUs \cite{kirk2007nvidia} are employed to enhance performance. FlashAttention \cite{dao2022flashattention, dao2023flashattention, shah2024flashattention, dao2023flashdecoding, hong2023flashdecoding++} related methods design I/O-aware exact attention speedup algorithms, leveraging GPU CUDA cores and on-chip SRAM to minimize access to High Bandwidth Memory (HBM), saving memory and reducing runtime. FuseMax \cite{nayak2024fusemax} uses Einsums and a spatial array accelerator, employing ping-pong scheduling to overlap MatMul and softmax operations.
    {While FlashAttention-3 \cite{shah2024flashattention} parallelizes MatMul and softmax on multi-core architectures, these cloud-based acceleration methods do not directly apply to resource-constrained edge accelerators, where there are limited number of processing units and on-chip memory.}

    To speed up attention inference on edge devices, current methods mainly leverage graph fusion \cite{ivanov2021data, niu2021dnnfusion, aminabadi2022deepspeed, mei2023defines} to restrict or reduce data transfers between off-chip and on-chip memory. 
    TVM \cite{chen2018tvm} utilizes an automated schedule optimizer to improve execution for a given neural network. Although TVM's auto-scheduler is designed for general purposes, the limitation is that it does not fuse MatMul and softmax operators in the attention block. 
    oneDNN \cite{li2024onednn} tackles the fusion of MatMul and softmax with graph fusion templates and microkernels to accelerate attention on Intel CPUs, while FLAT \cite{kao2023flat} uses row-granularity tiling and on-chip cache to alleviate the bandwidth bottleneck to access off-chip memory, achieving speedup and energy savings. Similarly, proprietary technologies like NVIDIA TensorRT \cite{NvidiaTensorRT} and Apple CoreML \cite{appleCoreML, appleMetalPSG, appleAccelerateF} claim to leverage graph fusion for attention acceleration. 

    
    Although these advancements promote fusion and data locality on edge devices, to the best of our knowledge, most existing works  in the public domain (e.g., FLAT) still execute the workloads including matrix multiplication (MatMul) and softmax sequentially, which achieves suboptimal latency. 
    It remains a significant challenge to execute heterogeneous workloads, including MatMul workloads that typically run on the multiplier accumulator (MAC) compute unit and softmax workloads that typically rely on the vector (VEC) unit, in parallel on edge accelerators with limited cores and thus limited or no multi-core parallelism opportunities.  
    Furthermore, the limited on-chip memory demands careful memory management schemes to prevent cache overflow and redundant computation. 

    In this paper, we introduce Memory Aware Stream Processing Attention (MAS-Attention) to accelerate attention computation on resource-constrained edge devices.
    MAS-Attention employs a semi-synchronous parallelization strategy to simultaneously utilize the heterogeneous MAC compute unit and vector compute unit on a neural accelerator in a pipelined parallel fashion, minimizing bubbles and optimizing the cachce management to improve attention inference efficiency. 
    Our contributions can be summarized as follows: 
    \begin{itemize}
    

        
        \item We propose a novel stream processing scheme that parallelizes tiled MatMul and softmax workloads through a semi-synchronous pipelining process. Prior works only parallelize the computing and I/O processes while still executing operators sequentially. In contrast, we aim to schedule all operators in the attention mechanism onto the heterogeneous computing units,  
        to achieve parallel execution by scheduling the stream of MatMul workloads on the MAC unit and stream of softmax workloads on the VEC unit while satisfying data dependencies between tiled workloads. 
        
        
        
        \item We employ a multi-tiered tiling scheme for MAS-Attention dataflow, that accommodates key hardware constraints and software parameters. This scheme employs fine-grained sub-matrix tiling for MatMul and row-granularity tiling for softmax operations. Using search strategies, we identify optimal tensor tiling factors to balance workloads efficiently within the stream processing scheme through offline auto-tuning across different attention workloads and hardware configurations.
        \item A proactive buffer overwrite strategy is further introduced to maintain efficiency with limited on-chip buffer capacity, especially for longer input sequences. This approach selectively overwrites specific MAC unit data to prioritize softmax completion with fine-grained control, minimizing data reloading. It ensures data dependencies, maintains operand integrity, and prevents pipeline stalls or reverting to prior rounds.

    \end{itemize}

    We extensively evaluate MAS-Attention across attention layers in transformer-based models, including different variants of BERT \cite{devlin2018bert}, Llama3-8B \cite{touvron2023llama}, T5 \cite{raffel2020exploring}, ViT \cite{dosovitskiy2020image} and XLM \cite{lample2019cross}. For simulations, we utilize a modified TileFlow \cite{zheng2023tileflow} to define the edge spatial accelerator architecture, software mapping, and search space exploration, while Timeloop \cite{parashar2019timeloop} and Accelergy \cite{wu2019accelergy} are used to estimate latency and energy consumption. Additionally, we test MAS-Attention on real hardware, using a Huawei MatePad Pro 13.2 with a DaVinci \cite{liao2019davinci} NPU. On the simulated edge device, MAS-Attention achieves up to $2.75×$ speedup and $54\%$ reduction in energy consumption compared to the state-of-the-art FLAT algorithm. Similar improvements in speedup and energy savings are also observed on the actual edge NPU hardware, further validating MAS-Attention's effectiveness.

%% file: sections/background.tex
\label{sec:background}

    \textbf{Sequential Attention Execution:}
    The Layer-Wise attention computation processes operations sequentially. This method relies on transferring intermediate results between off-chip and on-chip memory, creating a memory-bound workflow that poses significant deployment challenges on edge devices with limited memory bandwidth.

    \textbf{Approximate Attention Acceleration Methods:}
    For approximate acceleration methods of transformer-based foundation models,  methods like palletization \cite{cho2021dkm, tabani2021improving, wang2020cluster, appleCoreML}, quantization \cite{liu2021post, lin2021fq, wang2022deep, li2022q, piao2022sensimix, yao2022zeroquant, li2023vit, yu2023boost}, pruning \cite{mao2021tprune, peng2021accelerating, yu2022unified, yu2022width}, and knowledge distillation \cite{sun2019patient, wang2020minilm, wang2020minilmv2, ganesh2021compressing, huang2024knowledge, gupta2024progressive} compress model size by reducing parameters or transferring knowledge from larger models, achieving memory efficiency and faster inference.

    \textbf{Exact Attention Acceleration Methods:}
    In cloud environments, exact acceleration methods \cite{dao2022flashattention, dao2023flashattention, shah2024flashattention, dao2023flashdecoding, hong2023flashdecoding++, patel2023splitwise} leverage parallel computation on multi-core architectures to speed up attention mechanism. For instance, FlashAttention and FlashAttention-2 optimize dataflow for attention computation on NVIDIA A100 GPUs by dividing query, key and value inputs into smaller tiles and loading them from high-bandwidth memory (HBM) to on-chip SRAM, reducing data movement for large intermediate outputs and exploiting GPU CUDA core parallelism. FlashAttention-3 further enhances parallelism using ping-pong scheduling to overlap MatMul and softmax operations within warp groups on NVIDIA H100 GPUs. 
    {FuseMax \cite{nayak2024fusemax} leverages Einsums to implement fused attention computation on a spatial array accelerator, overlapping MatMul and softmax operations to enhance spatial PE array utilization.}

    Due to limited computing cores, resource-constrained edge devices rely on graph-fusion-based kernels \cite{gao1993collective, kjolstad2017tensor, chen2018learning, baghdadi2019tiramisu, googleXLA, zhou2022exploring}—such as oneDNN \cite{li2024onednn} for CPUs and FLAT \cite{kao2023flat}—to accelerate attention computations by fusing operators and retaining intermediate results on-chip, which reduces DRAM and off-chip memory access overhead. FLAT employs a row-based attention fusion strategy for TPUs \cite{jouppi2017datacenter, jouppi2020domain} and spatial accelerators \cite{kwon2018maeri, chen2019eyeriss}, including edge devices. By loading rows of query into on-chip memory, FLAT performs corresponding MatMul and softmax row-wise computations on-chip and writes the output rows directly to off-chip memory, thus mitigating memory-bound limitations by minimizing large data transfers. However, previous attention acceleration methods overlook the heterogeneous computing characteristics between MatMul and softmax, which run on MAC and VEC units, missing an opportunity for parallelization that could further reduce latency and energy consumption.

%% file: sections/challenge_motivation.tex
\label{sec:challenge_motivations}
    While prior exact attention acceleration methods for resource-constrained edge devices, such as oneDNN \cite{li2024onednn} and FLAT \cite{kao2023flat}, enhance data locality and reduce memory access overhead through operator fusion, they still execute tiled MatMul and softmax operators sequentially, missing the chance for parallel execution within the attention mechanism. In our work, we leverage the heterogeneous computing capabilities of edge devices to achieve parallel execution of tiled MatMul and softmax for the exact attention acceleration. Our method further minimizes the latency with this parallelization scheme while reducing I/O and redundant memory access with a novel multi-tiered tiling scheme and a proactive memory-aware buffer management, making it advantageous even for single inference requests in AI scenarios on resource-constrained edge devices.
    
    

    \textbf{Heterogeneous Workloads of Attention Mechanism:} 
    On resource-constrained edge devices, heterogeneous computing is often used to perform computation and memory access concurrently. However, prior edge-based attention acceleration methods have not explored the heterogeneous nature of MatMul and softmax workloads within the attention mechanism. Given their distinct computational characteristics, the compute-intensive MatMul operation runs on the MAC unit, while the element-wise softmax operation is processed on the VEC unit. Leveraging this heterogeneity, enables parallel execution of MatMul and softmax computations, providing further acceleration of the attention mechanism. 

    \textbf{Hardware-Software Co-design Scheduling on Resource-Constrained Edge Devices:} 
    Given the limited computing cores and on-chip memory, efficiently scheduling tiled MatMul and softmax operators with parallel execution in the attention workload requires consideration of both hardware parameters (e.g., L1 and L0 memory sizes, MAC and VEC core counts) and software parameters (e.g., MatMul and softmax workload shapes). To address this challenging hardware-software co-design scheduling problem, we propose a novel multi-tiered tiling scheme that accommodates both short and long sequence lengths while enhancing the utilization of on-chip processing units. Specifically, we introduce sub-matrix tiling granularity for MatMul and row-wise tiling granularity for softmax workloads. This approach creates distinct tiling search spaces for different workloads, allowing for higher search efficiency. We employ advanced search algorithms like MCTS to conduct offline searches for obtaining optimal tiling parameters across various attention workloads and hardware configurations. Our tiling scheme and search algorithm aim to balance MAC and VEC operations in a fused, pipelined, semi-synchronous attention computation, maximizing processing unit utilization, minimizing idle time, and reducing I/O and redundant memory access to ultimately optimize inference latency and energy consumption. 
    
    
    \textbf{Memory-aware Optimizations for the Limited Shared On-chip Memory:} 
    While our multi-tiered tiling scheme allocates search budgets for different workloads within the attention mechanism and enhances the efficiency of the search algorithm, limited search budgets can lead to locally optimal tiling parameters, particularly for long input sequences with extensive search spaces. Additionally, the constrained shared on-chip memory in edge devices complicates the scheduling of parallelized MatMul and softmax workloads. To address the potential for sub-optimal tiling parameters and better utilize limited on-chip memory, we introduce an innovative proactive buffer overwrite strategy. This memory-aware optimization features guardian mechanisms that proactively overwrite selected on-chip buffered data, balancing data refetching and redundant computation against cache overflow. It prioritizes critical operators for timely completion while ensuring correct data dependencies within the pipelined dataflow.

    \begin{figure*}[t]
      \centering
      \includegraphics[width=\textwidth]{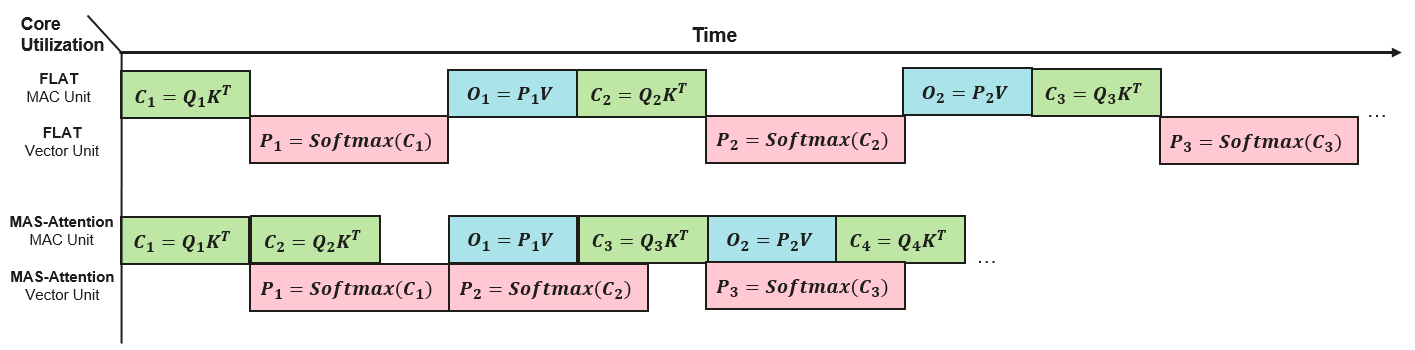}
      \vspace{-1.5em}
      \caption{Dataflow comparison between FLAT and MAS-Attention: FLAT executes tiled stages sequentially, while MAS-Attention performs MatMul and softmax operations semi-synchronously in parallel, maximizing compute utilization and significantly enhancing overall performance.}
      \label{fig:MAS_dataflow}
    \end{figure*}

%% file: sections/methodology.tex
Given the query, key and value matrices, $Q, K, V \in \mathbb{R}^{{B}\times{H}\times{N}\times{E}}$, where $B$ is the batch size, $H$ is the number of heads, $N$ is the sequence length and $E$ is the embedding size, the attention output $O$ is computed through the following steps: 
    \begin{align}
    \label{eq:attention_eq}
    & C = Q{K^T} \in \mathbb{R}^{{B}\times{H}\times{N}\times{N}},\\
    &
    \label{eq:attention_eq1}
    P = \mbox{softmax}(C) \in \mathbb{R}^{{B}\times{H}\times{N}\times{N}},\\
    &
    \label{eq:attention_eq2}
    O = PV \in \mathbb{R}^{{B}\times{H}\times{N}\times{E}},
    \end{align}
    where softmax is applied to every row of $QK^T$.

To efficiently perform these computations on resource-limited spatial accelerators, we propose a semi-synchronous MAC-VEC parallel execution scheme. Our method is applicable to a wide range of spatial accelerators that have at least one MAC unit for matrix multiplications and one VEC unit for element-wise operations. Our scheme is achieved through the strategic scheduling and pipelining of two MatMul operations alongside a single Softmax operation, as illustrated in \autoref{fig:MAS_dataflow}. This approach allows the three operators to concurrently process different tiles within the same computation round, thereby accelerating the attention mechanism. 
Additionally, we leverage advanced heuristic search algorithms to optimize the tiling sizes across all memory levels within our dataflow. These algorithms adaptively tune the tiling parameters based on input dimensions, workload characteristics, and pipelining criteria to ensure a balanced distribution of workloads across compute units.
We also implement an on-chip memory management strategy that selectively overwrites non-essential data to free up memory resources, prioritizing Softmax computation for longer sequences while ensuring the subsequent recovery of interrupted MatMul operations.
Detailed descriptions of these strategies are provided in the following. 


\subsection{Stream Processing Mechanism}

We propose a stream processing scheme to handle continuous streams of tiled MatMul and Softmax workloads. There are two streams of tiled tasks: one for tiled MatMul computation (defined in Algorithms \ref{alg:produce_C_i} and \ref{alg:produce_O_i}) and another for tiled Softmax computation defined in Algorithm \ref{alg:produce_P_i}. These streams are scheduled in a pipelined fashion to overlap tiled MatMul-Softmax computations, as illustrated in \autoref{fig:MAS_dataflow}.

Our approach operates at a row granularity, where the input matrix $Q$ is divided into smaller chunks along the batch, head, and sequence dimensions, resulting in row-wise sub-matrices denoted as $Q_i$. This granularity is driven by the inherently row-wise nature of the Softmax operation, aligning the processing scheme with Softmax’s requirements. Iterations thus proceed based on the segmented sequence dimension of the query, allowing for efficient parallelism. The detailed stream processing scheme is outlined in Algorithm \ref{alg:pipeline-attention-mechanism}, where there are \textit{warm-up}, \textit{regular}, and \textit{finalize} computation rounds.

\begin{algorithm}[ht]
\caption{MAS-Attention}
\label{alg:pipeline-attention-mechanism}
\begin{algorithmic}[1]
\STATE \textbf{Require:} $\mathbf{Q, K, V} \in \mathbb{R}^{B \times H \times N \times E}$ in DRAM; Parameters $B_b, H_h, N_Q, N_{K,V} \in \mathbb{R}$
\STATE Divide $\mathbf{Q}$ into $T_r = \left\lceil\frac{B}{B_b}\right\rceil \times \left\lceil\frac{H}{H_h}\right\rceil \times \left\lceil\frac{N}{N_Q}\right\rceil$ blocks $\mathbf{Q_1, ..., Q_r} \in \mathbb{R}^{B_b \times H_h \times N_Q \times E}$
\STATE Divide $\mathbf{O}$ into $T_r = \left\lceil\frac{B}{B_b}\right\rceil \times \left\lceil\frac{H}{H_h}\right\rceil \times \left\lceil\frac{N}{N_Q}\right\rceil$ blocks $\mathbf{O_1, ..., O_r} \in \mathbb{R}^{B_b \times H_h \times N_Q \times E}$
\STATE \textbf{Allocate} $(B, H, N, E)$ for $\mathbf{O}$ in DRAM
\STATE Call Alg. \ref{alg:produce_C_i}: $\mathbf{C_1} \gets \mathbf{Q_1 K^T}$
\STATE $i \gets 2$
\STATE \textbf{while} $i \leq T_r$ \textbf{do}
    \STATE \quad \textbf{if} $i = 2$ \textbf{then}
        \STATE \quad \quad \textbf{Parallel Execution:}
        \STATE \quad \quad \quad Call Alg. \ref{alg:produce_C_i}: $\mathbf{C_2} \gets \mathbf{Q_2 K^T}$ 
        \STATE \quad \quad \quad Call Alg. \ref{alg:produce_P_i}: $\mathbf{P_1} \gets \text{Softmax}(\mathbf{C_1})$ 
    \STATE \quad \textbf{else}
        \STATE \quad \quad \textbf{Parallel Execution:}
        \STATE \quad \quad \quad Call Alg. \ref{alg:produce_O_i}: $\mathbf{O_{i-2}} \gets \mathbf{P_{i-2} V}$ 
        \STATE \quad \quad \quad Call Alg. \ref{alg:produce_P_i}: $\mathbf{P_{i-1}} \gets \text{Softmax}(\mathbf{C_{i-1}})$ 
        \STATE \quad \quad \quad \textbf{Wait for completion of} Alg. \ref{alg:produce_O_i} \textbf{then:}
        \STATE \quad \quad \quad \quad Call Alg. \ref{alg:produce_C_i}: $\mathbf{C_i} \gets \mathbf{Q_i K^T}$ 
    \STATE \quad \textbf{end if}
    \STATE \quad $i \gets i + 1$
\STATE \textbf{end while}
\STATE \textbf{Finalize:}
\STATE \quad \textbf{Parallel Execution:}
\STATE \quad \quad Call Alg. \ref{alg:produce_O_i}: $\mathbf{O_{i-2}} \gets \mathbf{P_{i-2} V}$ 
\STATE \quad \quad Call Alg. \ref{alg:produce_P_i}: $\mathbf{P_{i-1}} \gets \text{Softmax}(\mathbf{C_{i-1}})$ 
\STATE \quad \textbf{Wait for completion of} Alg. \ref{alg:produce_P_i} \textbf{then:}
\STATE \quad \quad Call Alg. \ref{alg:produce_O_i}: $\mathbf{O_{i-1}} \gets \mathbf{P_{i-1} V}$ 
\STATE \textbf{return} $\mathbf{O}$
\end{algorithmic}
\end{algorithm}

In the \textit{warm-up} computation round, we use the MAC unit to compute the first tile for the first MatMul operator as $C_1 = Q_1K^T$. Then, we use the VEC unit to compute the first tile for the Softmax operator as $P_1 = \mbox{Softmax}(C_1)$ and use the MAC unit to compute the second tile for the first MatMul operator as $C_2 = Q_2K^T$ in parallel. 
Then we enter the \textit{regular} computation rounds, as shown in lines 13-17 of Algorithm \ref{alg:pipeline-attention-mechanism}. For iterations $i \geq 3$, the MAC unit computes the tile for the final MatMul operator as $O_{i-2} = P_{i-2}V$. Meanwhile, the VEC unit computes the tile for the Softmax operator as $P_{i-1} = \mbox{Softmax}(C_{i-1})$. While the tiled Softmax task is being processed, the MAC unit computes the tile for the first MatMul operator as $C_i = Q_iK^T$ upon completion of $O_{i-2}$. 
Lastly, in the \textit{finalize} computation round, the MAC unit computes the last tile for the final MatMul operator as $O_{i-1} = P_{i-1}V$ after the VEC unit computes the last tile of the Softmax operator as $P_{i-1} = \mbox{Softmax}(C_{i-1})$. 

Our pipelined attention mechanism operates in a semi-synchronous manner. During a \textit{regular} computation round, there is no data dependency among workloads, allowing the two tiled MatMuls and Softmax to be executed in parallel by the MAC and VEC units, respectively. However, within each computation round, data dependencies must be carefully managed to ensure the correctness of the computation. This semi-synchronous MAC-VEC parallelism for MatMul-Softmax computations significantly reduces the latency of the attention mechanism. 

\subsection{MAS-Attention Tiling Scheme}

We introduce a multi-tiered tiling strategy for MAS-Attention dataflow. For matrices $K$, $P$ and $V$, used in the MatMul operations in \autoref{eq:attention_eq} and \autoref{eq:attention_eq2}, a fine-grained sub-matrix tiling is applied. This approach is crucial, especially when the sequence length is significantly longer than the embedding dimension ($N \gg E$), as it helps address the constraints of limited on-chip memory. Without such tiling, handling the matrix $K$ in $C_i = Q_iK^T$ and the matrices $P_i$ and $V$ in $O_i = P_iV$ becomes problematic due to excessive memory demands. For intermediate tensors $C_i$ and $P_i$ used in the Softmax operation in \autoref{eq:attention_eq1}, a row-granularity tiling is employed, aligning with the inherent row-wise nature of Softmax to maintain computational correctness.

We establish a comprehensive search space for tiling parameters across the memory hierarchy of the targeted hardware, focusing on dimensions such as batch size ($B$), number of attention heads ($H$), query sequence length ($N_Q$), and key/value sequence lengths ($N_{K,V}$). The search for optimal tiling parameters is influenced by three key factors: the detailed workload of attention mechanism, the specific scheduling of MAS-Attention, and the input size. These parameters are defined at each memory level to ensure efficient off-chip and on-chip memory operations while considering the interaction between computation and memory usage. This approach aims to identify optimal or near-optimal tiling configurations that maintain computational efficiency throughout the stream processing of MAS-Attention. 
To effectively navigate this search space, we use Genetic Algorithms and Monte Carlo Tree Search (MCTS) for the simulated edge device, and Grid Search for the edge device with a DaVinci DNN Accelerator.

{We use MCTS to optimize tiling factors. At each step, MCTS selects a loop and assigns a tiling factor based on the number of iterations the loop will execute, updating constraints and passing them to the next untiled loop. Once all tiling factors are determined, a complete fusion mapping is produced as an analysis tree where each node corresponds to a tile, which is then evaluated. The results of each evaluation are fed back to MCTS to update the upper confidence bounds (UCB), guiding subsequent searches.
Genetic algorithm (GA) then aims to find optimal compute ordering in the analysis tree based on the found tiling factors, refining performance across different analysis trees. GA generates a population of analysis trees, applies crossover and mutation, and evaluates each tree using the tiling factors. Through repeated iterations, the best analysis tree is selected as the optimal fusion dataflow.}


On the DaVinci DNN Accelerator, Grid Search systematically evaluates all possible configurations, leveraging its compatibility with the hardware's structured memory model. These algorithms iteratively assess various tiling configurations, simulating different tile shapes and sizes to determine those that optimize execution cycles and minimize power consumption.

After retrieving the optimal tiling parameters from the search, Algorithm \autoref{alg:produce_C_i} performs the tiled MatMul computation of $C_i$, using sub-matrices $Q_i$ and finer-grained sub-tiles of $K$. The algorithm reads blocks of $Q_i$ and sub-blocks of $K$ from DRAM to on-chip memory, where the MatMul operation $Q_i K^T$ is executed to generate $C_i$. The resulting $C_i$ is retained on-chip for subsequent operations.

\begin{algorithm}[ht]
\caption{Produce $C_i \gets Q_iK^T$}
\label{alg:produce_C_i}
\begin{algorithmic}[1]
\STATE \textbf{Require:} $\mathbf{Q_i} \in \mathbb{R}^{B_b \times H_h \times N_Q \times E}$, $\mathbf{K} \in \mathbb{R}^{B \times H \times N \times E}$ in DRAM; $B_b, H_h, N_Q, N_{K,V} \in \mathbb{R}$; $i = [b_b:b_e, h_b:h_e, n_b:n_e]$
\STATE Select $i^{th}$ set of batch and head from $\mathbf{K}$ as $\mathbf{K_i}=\mathbf{K}[i[0], i[1],:,:] \in \mathbb{R}^{B_b \times H_h \times N \times E}$
\STATE Divide $\mathbf{K_i}$ into $T_c = \left\lceil \frac{N}{N_{K,V}} \right\rceil$ blocks $\mathbf{K_{i,1}}, \ldots, \mathbf{K_{i,c}} \in \mathbb{R}^{B_b \times H_h \times N_{K,V} \times E}$
\STATE Allocate $(B_b, H_h, N_Q, N)$ for $\mathbf{C_i}$, $(B_b, H_h, N_Q, E)$ for $\mathbf{Q_i}$, and $(B_b, H_h, N_{K,V}, E)$ for $\mathbf{K_{i,j}}$ in on-chip memory
\STATE Load $\mathbf{Q_i}$ from DRAM to on-chip memory
\FOR{$1 \leq j \leq T_c$}
    \STATE Load $\mathbf{K_{i,j}}$ from DRAM to on-chip memory
    \STATE On-chip compute $\mathbf{C_{i,j}} = \mathbf{Q_i K_{i,j}^T} \in \mathbb{R}^{B_b \times H_h \times N_Q \times N_{K,V}}$\\
    \STATE Write $\mathbf{C_{i,j}}$ to on-chip memory as $j^{th}$ block of $\mathbf{C_i}$
\ENDFOR
\end{algorithmic}
\end{algorithm}


Algorithm \ref{alg:produce_P_i} handles the tiled softmax computation. It processes the on-chip $C_i$ matrix by dividing it into smaller row-wise blocks, aligned with the row-wise nature of the softmax operation. Each block undergoes the softmax steps: identifying the maximum value, subtracting it, exponentiating, summing, and normalizing to produce $P_i$. The $P_i$ blocks are kept on-chip to ensure efficient data access for final Matmul computation.


\begin{algorithm}[ht]
\caption{Produce $P_i \gets C_i$}
\label{alg:produce_P_i}
\begin{algorithmic}[1]
\STATE \textbf{Require:} $\mathbf{C_i} \in \mathbb{R}^{B_b \times H_h \times N_Q \times N}$ in on-chip memory; $B_b, H_h, N_Q \in \mathbb{R}$; $i = [b_b:b_e, h_b:h_e, n_b:n_e]$
\STATE Divide $\mathbf{C_i}$ into $T_l = N_Q$ blocks $\mathbf{C_{i,1}}, \ldots, \mathbf{C_{i,l}} \in \mathbb{R}^{B_b \times H_h \times 1 \times N}$
\STATE Allocate $(B_b, H_h, N_Q, N)$ for $\mathbf{P_i}$ in on-chip memory
\FOR{$1 \leq j \leq T_l$}
    \STATE On-chip compute $P_{i,j} = \text{Softmax}(\mathbf{C_{i,j}}) \in \mathbb{R}^{B_b \times H_h \times 1 \times N}$
    \STATE Write $\mathbf{P_{i,j}}$ to on-chip memory as $j^{th}$ block of $\mathbf{P_i}$
\ENDFOR
\end{algorithmic}
\end{algorithm}

Algorithm \ref{alg:produce_O_i} handles the tiled MatMul computation of $O_i$. Both $P_i$ and $V_i$ are divided into finer-grained blocks to manage large sequence lengths. In each iteration, a block of $V_i$ is loaded from DRAM to on-chip memory, while a corresponding block of $P_i$ is already available on-chip. The block-wise multiplication $P_{i,j} V_{i,j}$ is performed iteratively, accumulating results into $O_i$. Once all iterations are complete, $O_i$ is written back to off-chip memory.


\begin{algorithm}[ht]
\caption{Produce $O_i \gets P_iV$}
\label{alg:produce_O_i}
\begin{algorithmic}[1]
\STATE \textbf{Require:} $\mathbf{P_i} \in \mathbb{R}^{B_b \times H_h \times N_Q \times N}$ in on-chip memory; $\mathbf{V} \in \mathbb{R}^{B \times H \times N \times E}$, $\mathbf{O} \in \mathbb{R}^{B \times H \times N \times E}$ in DRAM; $B_b, H_h, N_Q, N_{K,V} \in \mathbb{R}$; $i = [b_b:b_e, h_b:h_e, n_b:n_e]$\\
\STATE Select $i^{th}$ set batch and head from $\mathbf{V}$ as $\mathbf{V_i}=\mathbf{V}[i[0], i[1],:,:] \in \mathbb{R}^{B_b \times H_h \times N \times E}$
\STATE Divide $\mathbf{V_i}$ into $T_c = \left\lceil \frac{N}{N_{K,V}} \right\rceil$ blocks $\mathbf{V_{i,1}}, \ldots, \mathbf{V_{i,c}} \in \mathbb{R}^{B_b \times H_h \times N_{K,V} \times E}$
\STATE Divide $\mathbf{P_i}$ into $T_c = \left\lceil \frac{N}{N_{K,V}} \right\rceil$ blocks $\mathbf{P_{i,1}}, \ldots, \mathbf{P_{i,c}} \in \mathbb{R}^{B_b \times H_h \times N_Q \times N_{K,V}}$
\STATE Allocate $(B_b, H_h, N_Q, E)$ for $\mathbf{O_i}$, and $(B_b, H_h, N_{KV}, E)$ for $\mathbf{V_{i,j}}$ in on-chip memory
\STATE On-chip initialize $\mathbf{O_i} = (0)_{B_b \times H_h \times N_Q \times E} \in \mathbb{R}^{B_b \times H_h \times N_Q \times E}$
\FOR{$1 \leq j \leq T_c$}
    \STATE Load $\mathbf{V_{i,j}}$ from DRAM to on-chip memory
    \STATE On-chip compute $\mathbf{O_i} = \mathbf{O_i} + \mathbf{P_{i,j} V_{i,j}} \in \mathbb{R}^{B_b \times H_h \times N_Q \times E}$
\ENDFOR
\STATE Write $\mathbf{O_i}$ to off-chip memory as $i^{th}$ block of $\mathbf{O}$
\end{algorithmic}
\end{algorithm}



\subsection{Proactive Overwrite Strategy for Optimized Memory Utilization} \label{sec:Memory-Efficient-Optimization}

The tiling parameters obtained from heuristic search algorithms, such as Genetic Algorithm and Monte Carlo Tree Search, may not always yield optimal results. Due to the complexity of the search space and the heuristic nature of these algorithms, there is a possibility of suboptimal configurations, which can impact the efficiency and correctness of stream processing. To mitigate these potential inefficiencies and ensure robust performance across a variety of workloads and scenarios, we introduce a selective overwrite strategy. This proactive approach enables the system to adaptively manage on-chip memory by selectively overwriting specific non-essential data when memory constraints arise. 

\begin{figure}[t]
  \centering
  \adjustbox{max width=\columnwidth, max height=0.5\textheight}{\includegraphics{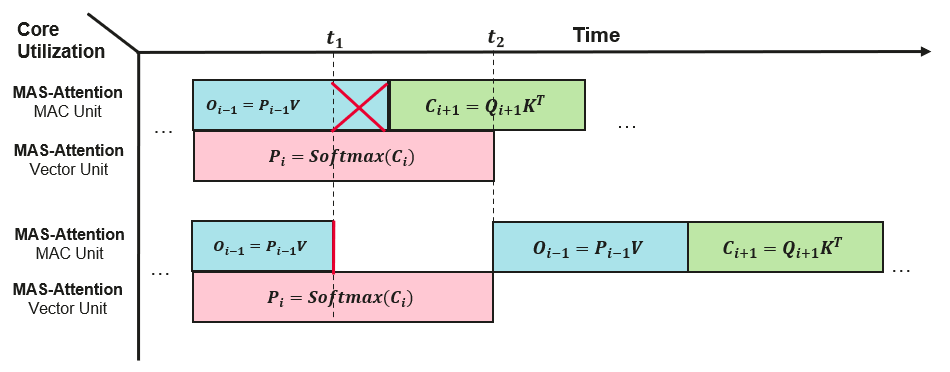}}
  \vspace{-1.5em}
  \caption{Selective Overwriting of $V$ Matrix to Halt MatMul Operation in MAS-Attention’s Memory Strategy.}
  \label{fig:overwrite_V}
\end{figure}

\begin{figure}[t]
  \centering
  \adjustbox{max width=\columnwidth, max height=0.5\textheight}{\includegraphics{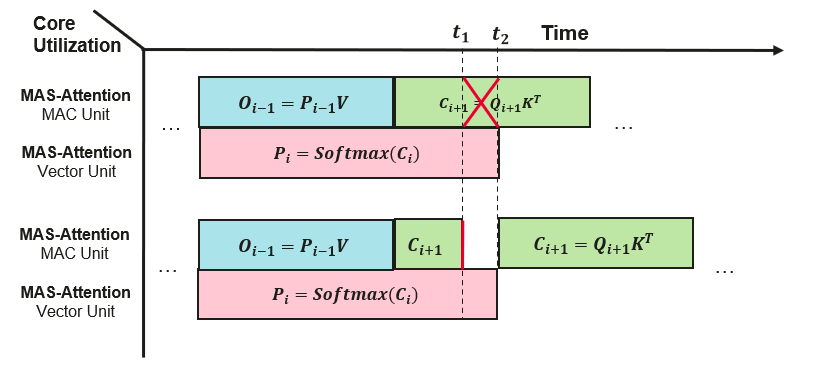}}
  \vspace{-1.5em}
  \caption{Selective Overwriting of $K$ Matrix to Halt MatMul Operation in MAS-Attention’s Memory Strategy.}
  \label{fig:overwrite_K}
\end{figure}

During the computation of $P_i$, if the on-chip memory reaches capacity, impeding further calculations, two cases may arise. First, as shown in \autoref{fig:overwrite_V}, if the MAC unit is engaged in processing $P_{i-1}V$, $P_i$ will overwrite the $V$ matrix on chip and stop the MAC from continuing its operation, resulting in no more writes from the MAC unit to on-chip buffer. 
Second, as shown in \autoref{fig:overwrite_K}, if the MAC unit is occupied with $Q_{i+1}K^T$, $P_i$ will overwrite the $K$ matrix on chip, thereby interrupting the MAC unit's process and preventing any further writes to the on-chip memory. Once the final result of $P_i$ is fully calculated and stored on chip, the MAC unit can resume its process by reloading either the $V$ or $K$ matrix from DRAM to on-chip memory if it was overwritten and redoing the MatMul calculation.

The rationale is that maintaining the integrity of critical operands is essential to the efficiency of the pipeline.
Preserving and finishing $P_i = \mbox{softmax}(C_i)$ is crucial as the softmax operation stores its results only on chip and depends on $C_i=Q_iK^T$ which was obtained from on-chip memory, hence overwriting $P_i$ cannot be remedied by reloading it from DRAM. In contrast, this is not the case for $K$ and $V$ matrices, overwriting of which can be remedied by reloading the over-written tensors from DRAM without stalling the pipeline computation rounds.

This strategy ensures efficient use of on-chip memory and computational resources, and our careful data overwrite method makes the impact of the increased number of DRAM reads on our overall latency and energy savings unnoticeable. By carefully managing memory overwrites and reloading only essential data, we strike a balance between maximizing parallelism and maintaining computational efficiency, ultimately leading to improved performance and energy efficiency.

%% file: sections/experiment_2.tex
\label{sec:experiment_2}



\subsection{Experimental Setup}
This section provides details on the simulation and modeling tools utilized, describes the hardware specifications, and outlines the experimental workloads and baseline algorithms used for analysis. 
We conduct a comprehensive evaluation of the proposed method, comparing it against state-of-the-art attention fusion and acceleration techniques tailored for spatial accelerators in edge environments. This includes an assessment of performance on foundation model workloads suited for edge deployment.

\begin{figure}[t]
  \centering
  \includegraphics[width=\columnwidth]{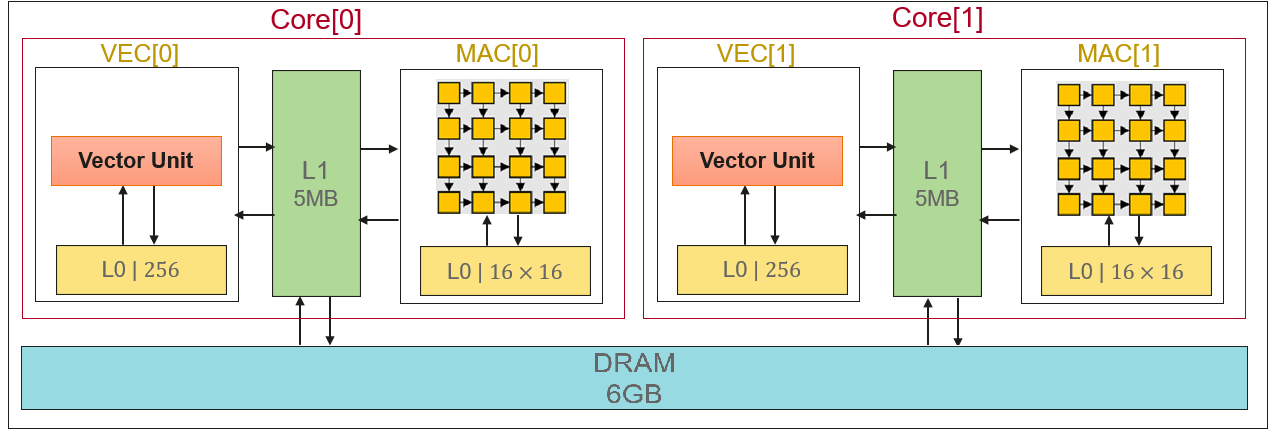}
  \vspace{-1.5em}
  \caption{Simulated Edge Architecture Design}
  \label{fig:System_Config}
\end{figure}

\textbf{Simulation and Modeling Tools:} {To simulate our experiments, we employed Timeloop \cite{parashar2019timeloop} and Accelergy \cite{wu2019accelergy} to measure the latency and energy consumption, also we modified TileFlow \cite{zheng2023tileflow} to define the edge spatial accelerator, software mapping for attention inference, and search space exploration. During the tiling and loop parameters search, MCTS generated tiling factors and GA refined compute orderings, with each candidate evaluated using Timeloop/Accelergy.}
The custom edge hardware architecture designed for simulation operates at a frequency of 3.75GHz and features 16nm technology, two cores each containing a MAC and a VEC unit, and a hierarchical memory system as depicted in \autoref{fig:System_Config}. The designed DRAM has a bandwidth of 30GB/sec and a total size of 6GB. The L1 cache has connection to DRAM and L0 register file and has a storage of 5MB . The Processing Elements (PEs) in MAC and VEC units, organized in 16x16 and 256 mesh respectively, have access to L0 register file. These parameters for the hardware architecture were determined after various stress tests of the hardware.
Our simulations were conducted on a system equipped with an Intel(R) Xeon(R) Gold 6140 CPU @ 2.30GHz, utilizing a single thread in execution.
Additionally, we evaluated our algorithm on a real hardware, Huawei MatePad Pro 13.2, to validate its practical applicability and performance. 
{More specifically, this device is equipped with the Kirin 990 5G SoC featuring Da Vinci NPU architecture, which consists of three cores—each with a MAC unit, a Vector Unit, and dedicated on-chip memory. The NPU includes 2x Ascend Lite cores and 1x Ascend Tiny core.} 

\textbf{Workloads:} The workload for our experiments focuses on the inference of attention layers in various transformer-based networks, including different variants of BERT \cite{devlin2018bert}, Llama \cite{touvron2023llama}, T5 \cite{raffel2020exploring}, ViT \cite{dosovitskiy2020image}, and XLM \cite{lample2019cross}, as detailed in \autoref{table:modelspecs}. 
{We also provide end-to-end results from deploying MAS-Attention in a real-world AI workload based on Stable Diffusion 1.5 UNet discussed further in \ref{subsubsec:Analysis_on_Real_Hardware}. 
We selected a diverse set of networks with varying attention layer dimensions to ensure a comprehensive evaluation.} Additionally, each workload in our experiments undergoes a rigorous golden data check for all methods, including our proposed approach, ensuring that all methods pass this validation.

\textbf{Layer-Wise:} This approach represents the unfused baseline for attention inference. In this method, $C = QK^T$ is fully computed first, followed by the the softmax function on the entire matrix $C$ to yield $P = \text{Softmax}(C)$. Once $P$ is fully computed, the final output of the attention unit, $O = PV$, is then calculated. All these operations occur sequentially and without fusion.

\textbf{Soft-Pipe:} For comparison, we also design a baseline algorithm that only pipelines the first MatMul and the softmax. It divides $Q$ and $K$ into smaller chunks, fuses and pipelines MatMul of $C = QK^T$ with $\text{Softmax}(C)$. In each iteration, rows of $Q$ ($Q_{i}$) are loaded into on-chip memory to compute the corresponding rows of $C$, where $C_i = Q_iK^{T}$. Then, the corresponding rows of $P$ are computed on-chip with $P_i = \text{Softmax}(C_i)$. While $P_i$ is being calculated, $C_{i+1}$ can be computed simultaneously. The resulting $P$ values are stored back to DRAM, and once the computation of $P$ is complete, the final output of the attention unit, $O = PV$, is calculated sequentially.

\textbf{FLAT:} $Q$, $K$, and $V$ matrices are divided into smaller chunks, and all attention operations are fused on-chip and computed sequentially. In each iteration, rows of $Q$ ($Q_i$) are loaded into on-chip memory to compute the corresponding rows of $C$, where $C_i = Q_iK^{T}$. Then, the corresponding rows of $P$ are computed on-chip with $P_i = \text{Softmax}(C_{i})$. Finally, the corresponding rows of $O$, where $O_i = P_iV$, are computed on-chip and written back to off-chip memory.



\textbf{TileFlow:} In this approach, $Q$, $K$, and $V$ are divided into smaller chunks, and all operations in the attention unit are fused and pipelined \cite{zheng2023tileflow}. However, since \cite{zheng2023tileflow} does not provide further implementation details, we implemented the algorithm to the best of our knowledge based on the available information. Specifically, we replicated TileFlow’s tiling and pipelining approach by dividing matrices into sub-tiles that fit within on-chip memory, fusing MatMul and softmax operations with pipeline execution. This implementation approximates TileFlow’s behavior, ensuring our evaluation aligns with its intended operational characteristics.


{\textbf{FuseMax (scaled down to edge device):} The computation is decomposed into a sequence of 12 primitive operators based on extended einsum notation, which are executed using pipelining. The attention scores are computed as $C = QK^T$, and the Softmax function is implemented through a series of sub-operations, where MAC and VEC are processed in parallel. The weighted sum with $V$ is fused into the Softmax pipeline itself. All computations are fused and executed in a single pass.}


\begin{table}[ht]
\centering
\caption{Network Configuration and Hyper-Parameters}
\label{table:modelspecs}
\renewcommand{\arraystretch}{1.4}
\scriptsize
\begin{tabular}{|c|c|c|c|c|}
\hline
\textbf{Network Name} & \textbf{\#Heads} & \textbf{\#Seq} & \textbf{Hidden size} & \textbf{Emb\textsubscript{K,V}} \\ \hline
BERT-Base \& T5-Base & 12 & 512 & 768 & 64 \\ \hline
BERT-Large \& T5-Large & 16 & 512 & 1024 & 64 \\ \hline
BERT-Small & 8 & 512 & 512 & 64 \\ \hline
Llama3-8B \& T5-3B (T5-XL) & 32 & 512 & 4096 & 128 \\ \hline
T5-Mini \& T5-Small & 8 & 512 & 256 & 32 \\ \hline
ViT-B/14 & 12 & 196 & 768 & 64 \\ \hline
ViT-L/14 & 16 & 196 & 1024 & 64 \\ \hline
ViT-H/14 & 16 & 196 & 1280 & 80 \\ \hline
ViT-B/16 & 12 & 256 & 768 & 64 \\ \hline
ViT-L/16 & 16 & 256 & 1024 & 64 \\ \hline
ViT-H/16 & 16 & 256 & 1280 & 80 \\ \hline
XLM & 8 & 512 & 1024 & 128 \\ \hline
\end{tabular}
\end{table}

\subsection{Execution Time Analysis}

\subsubsection{Analysis on Simulated Hardware}
Table \ref{tab:cycles_speedup} presents a detailed analysis of execution cycles and speedup ratios for MAS-Attention compared to other methods across all tested networks. {The data highlights that MAS-Attention consistently achieves superior performance, with speedup factors up to $8.50\times$ over Layer-Wise, $4.5\times$ over Soft-Pipe, $2.75\times$ over FLAT, $1.75\times$ over TileFlow, and $1.47\times$ over FuseMax methods. The geometric means of these speedup values—$5.09\times$, $2.78\times$, $1.70\times$, $1.31\times$, and $1.27\times$ respectively—demonstrate MAS-Attention's overall efficiency in reducing execution time.} This substantial performance improvement underscores MAS-Attention's effectiveness as an advanced solution for optimizing computational efficiency in attention mechanisms.

\begin{table*}[ht]
    \centering
    \caption{Cycles and Speedup Comparisons Across Networks for Different Methods}
    \label{tab:cycles_speedup}
    \small
    \resizebox{\textwidth}{!}{
    \begin{tabular}{|c|c|c|c|c|c|c|c|c|c|c|c|}
        \hline
        \multirow{2}{*}{\textbf{Network Name}} & \multicolumn{6}{c|}{\textbf{Cycles }($10^6$)} & \multicolumn{5}{c|}{\textbf{Speedup (MAS-Attention vs. Others)}} \\ \cline{2-12}
         & Layer-Wise & Soft-Pipe & FLAT & TileFlow & FuseMax & MAS-Attention & Layer-Wise & Soft-Pipe & FLAT & TileFlow & FuseMax \\ \hline
        BERT-Base \& T5-Base & 3.637 & 2.064 & 1.573 & 0.799 & 0.992 & 0.786 & 4.63 & 2.63 & 2.00 & 1.02 & 1.26 \\ \hline
        BERT-Large \& T5-Large & 5.505 & 2.753 & 1.835 & 1.311 & 1.323 & 1.049 & 5.25 & 2.63 & 1.75 & 1.25 & 1.26 \\ \hline
        BERT-Small & 2.753 & 1.376 & 0.918 & 0.655 & 0.661 & 0.524 & 5.25 & 2.63 & 1.75 & 1.25 & 1.26 \\ \hline
        Llama3-8B \& T5-3B (T5-XL) & 12.845 & 8.389 & 4.719 & 5.243 & 4.864 & 4.194 & 3.06 & 2.00 & 1.13 & 1.25 & 1.16 \\ \hline
        T5-Mini \& T5-Small & 2.228 & 1.180 & 0.721 & 0.328 & 0.384 & 0.262 & 8.50 & 4.50 & 2.75 & 1.25 & 1.47  \\ \hline
        ViT-B/14 & 0.612 & 0.381 & 0.266 & 0.263 & 0.196 & 0.151 & 4.06 & 2.53 & 1.77 & 1.75 & 1.30 \\ \hline
        ViT-L/14 & 1.242 & 0.508 & 0.354 & 0.351 & 0.262 & 0.201 & 6.19 & 2.53 & 1.77 & 1.75 & 1.30 \\ \hline
        ViT-H/14 & 1.355 & 0.558 & 0.405 & 0.439 & 0.318 & 0.251 & 5.40 & 2.23 & 1.61 & 1.75 & 1.27 \\ \hline
        ViT-B/16 & 1.081 & 0.590 & 0.426 & 0.249 & 0.259 & 0.197 & 5.50 & 3.00 & 2.17 & 1.27 & 1.32 \\ \hline
        ViT-L/16 & 1.311 & 0.786 & 0.524 & 0.332 & 0.346 & 0.262 & 5.00 & 3.00 & 2.00 & 1.27 & 1.32 \\ \hline
        ViT-H/16 & 1.376 & 0.852 & 0.590 & 0.414 & 0.419 & 0.328 & 4.20 & 2.60 & 1.80 & 1.26 & 1.28 \\ \hline
        XLM & 4.194 & 2.097 & 1.180 & 1.311 & 1.216 & 1.049 & 4.00 & 2.00 & 1.13 & 1.25 & 1.16 \\ \hline
        \hline
        \textbf{Geometric Mean} & - & - & - & - & - & - & \textbf{5.09x} & \textbf{2.78x} & \textbf{1.70x} & \textbf{1.31x} & \textbf{1.27x} \\ \hline
    \end{tabular}
    }
\end{table*}

\subsubsection{Analysis on Real Hardware} \label{subsubsec:Analysis_on_Real_Hardware}
\begin{figure}[t]
  \centering
  \includegraphics[width=\columnwidth]{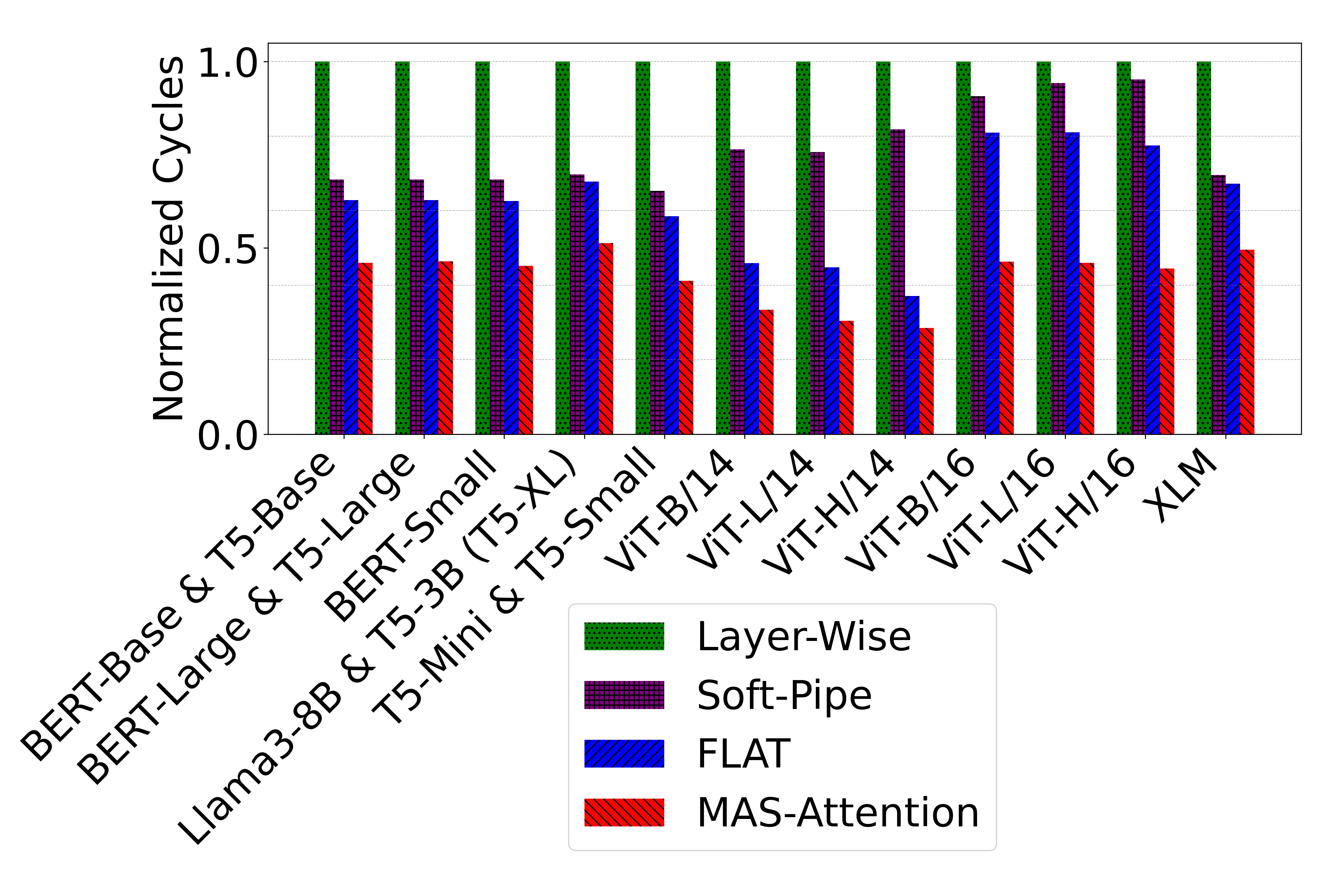}
  \vspace{-1.5em}
  \caption{Normalized Execution Time Comparison Across Networks for Different Methods on Huawei MatePad Pro 13.2 with DaVinci DNN Accelerator}
  \label{fig:CPRO_cycleAnalysis}
\end{figure}

\autoref{fig:CPRO_cycleAnalysis} shows the analysis of normalized execution time for Layer-Wise, Soft-Pipe, FLAT, and MAS-Attention methods on Huawei MatePad Pro 13.2 with DaVinci DNN Accelerator. MAS-Attention achieves substantial performance improvements, with speedups ranging from $1.94\times$ to $3.50\times$ over Layer-Wise, $1.35\times$ to $2.87\times$ over Soft-Pipe, and $1.30\times$ to $1.76\times$ over FLAT. The geometric mean speedups are $2.33\times$, $1.73\times$, and $1.42\times$, respectively. It is worth noting that TileFlow was not included in this analysis as its implementation details were not fully described in \cite{zheng2023tileflow}, which limited us from deploying it on this edge device. Overall, the data validates MAS-Attention's effectiveness in enhancing computational efficiency on real hardware.

{Additionally, to provide end-to-end experimental results, we evaluated MAS-Attention on a real-world generative AI workload, specifically a reduced UNet module of Stable Diffusion 1.5 running directly on the mobile device. This UNet contains 15 attention units, with the largest attention layer featuring 2 heads, a sequence length of 4096, and an embedding size of 64. Compared to the Layer-Wise method, MAS-Attention achieved a 29.4\% runtime reduction for the largest attention unit and a 6\% overall reduction in end-to-end model inference latency, further demonstrating the practical effectiveness of our proposed algorithm.}

\subsection{Power and Energy Analysis}


Table \ref{tab:energy_savings} presents a comprehensive analysis of energy consumption and savings achieved by MAS-Attention compared to other methods across various networks. The data reveals that MAS-Attention consistently demonstrates significant energy consumption reductions over Layer-Wise, Soft-Pipe, FLAT, and TileFlow, with savings ranging from $39.16\%$ to $66.67\%$, $39.61\%$ to $75.00\%$, $0.02\%$ to $54.03\%$, and $36.83\%$ to $65.05\%$, respectively. The geometric mean of these savings—$52.97\%$, $63.07\%$, $18.55\%$, and $53.16\%$—highlights MAS-Attention's overall effectiveness in reducing energy consumption.
{When compared to FuseMax, MAS-Attention achieves lower energy consumption for ViT-B/14, ViT-L/14, ViT-H/14, ViT-L/16, and ViT-H/16 but exhibits higher energy usage in other cases. The reason is that our objective in the search framework was to minimize latency rather than energy, although MAS-Attention can be revised to optimize other objectives. Nevertheless, MAS-Attention remains competitive in these results by maintaining a strong balance between energy efficiency and overall computational cycles.}

\begin{table*}[ht]
    \centering
    \caption{Energy Consumption and Savings Comparisons Across Networks for Different Methods.}
    \label{tab:energy_savings}
    \small
    \resizebox{\textwidth}{!}{
    \begin{tabular}{|c|c|c|c|c|c|c|c|c|c|c|c|}
        \hline
        \multirow{2}{*}{\textbf{Network Name}} & \multicolumn{6}{c|}{\textbf{Energy Consumption} ($10^9$ pJ)} & \multicolumn{5}{c|}{\textbf{Energy Savings (MAS-Attention vs. Others)}} \\ \cline{2-12}
         & Layer-Wise & Soft-Pipe & FLAT & TileFlow & FuseMax & MAS-Attention & Layer-Wise & Soft-Pipe & FLAT & TileFlow & FuseMax \\ \hline
        BERT-base \& T5-Base & 37.208 & 49.607 & 12.656 & 27.598 & 10.217 & 12.405 & 66.67\% & 75.00\% & 1.98\% & 55.05\% & {-21.42\%} \\ \hline
        BERT-large \& T5-Large & 28.105 & 65.672 & 21.112 & 38.065 & 13.623 & 16.944 & 39.69\% & 74.20\% & 19.75\% & 55.49\% & {-24.38\%} \\ \hline
        BERT-small & 20.218 & 24.336 & 10.556 & 19.032 & 6.811 & 8.359 & 58.65\% & 65.64\% & 20.80\% & 56.08\% & {-22.73\%} \\ \hline
        Llama3-8B \& T5-3B (T5-XL) & 179.309 & 186.463 & 63.252 & 147.502 & 53.401 & 63.241 & 64.73\% & 66.08\% & 0.02\% & 57.12\% & {-18.43\%} \\ \hline
        T5-Mini \& T5-Small & 12.434 & 11.269 & 8.744 & 7.512 & 3.542 & 4.746 & 61.83\% & 57.90\% & 45.71\% & 36.83\% & {-33.99\%} \\ \hline
        ViT-B/14 & 3.720 & 7.376 & 2.803 & 4.136 & 2.104 & 1.903 & 48.87\% & 74.21\% & 32.11\% & 54.00\% & {9.56\%} \\ \hline
        ViT-L/14 & 5.539 & 7.335 & 5.648 & 7.428 & 2.805 & 2.596 & 53.13\% & 64.61\% & 54.03\% & 65.05\% & {7.45\%} \\ \hline
        ViT-H/14 & 6.585 & 9.120 & 4.741 & 6.783 & 3.487 & 3.162 & 51.98\% & 65.34\% & 33.27\% & 53.38\% & {9.31\%} \\ \hline
        ViT-B/16 & 5.323 & 5.828 & 3.350 & 7.119 & 3.187 & 3.239 & 39.16\% & 44.42\% & 3.34\% & 54.49\% & {-1.63\%} \\ \hline
        ViT-L/16 & 9.403 & 6.984 & 6.316 & 9.402 & 4.249 & 4.218 & 55.14\% & 39.61\% & 33.21\% & 55.14\% & {0.73\%} \\ \hline
        ViT-H/16 & 11.160 & 15.414 & 6.803 & 11.475 & 5.278 & 5.156 & 53.81\% & 66.55\% & 24.22\% & 55.09\% & {2.31\%} \\ \hline
        XLM-Base & 35.786 & 46.485 & 15.813 & 36.876 & 13.350 & 15.584 & 56.45\% & 66.47\% & 1.45\% & 57.74\% & {-16.77\%} \\ \hline
        \hline
        \textbf{Geometric Mean} & - & - & - & - & - & - & \textbf{52.97\%} & \textbf{63.07\%} & \textbf{18.55\%} & \textbf{53.16\%} & \textbf{-11.94\%} \\ \hline
    \end{tabular}
    }
    \parbox{\textwidth}{
        \centering
        \vspace{0.5em}
        \footnotesize{Note: Based on some literature studies, ``pJ'' (picojoule) is used as the unit for energy consumption reported by Accelergy. Negative savings indicate higher energy consumption compared to the baseline.}
    }
\end{table*}

\begin{figure}[ht]
  \centering
  \includegraphics[width=\columnwidth]{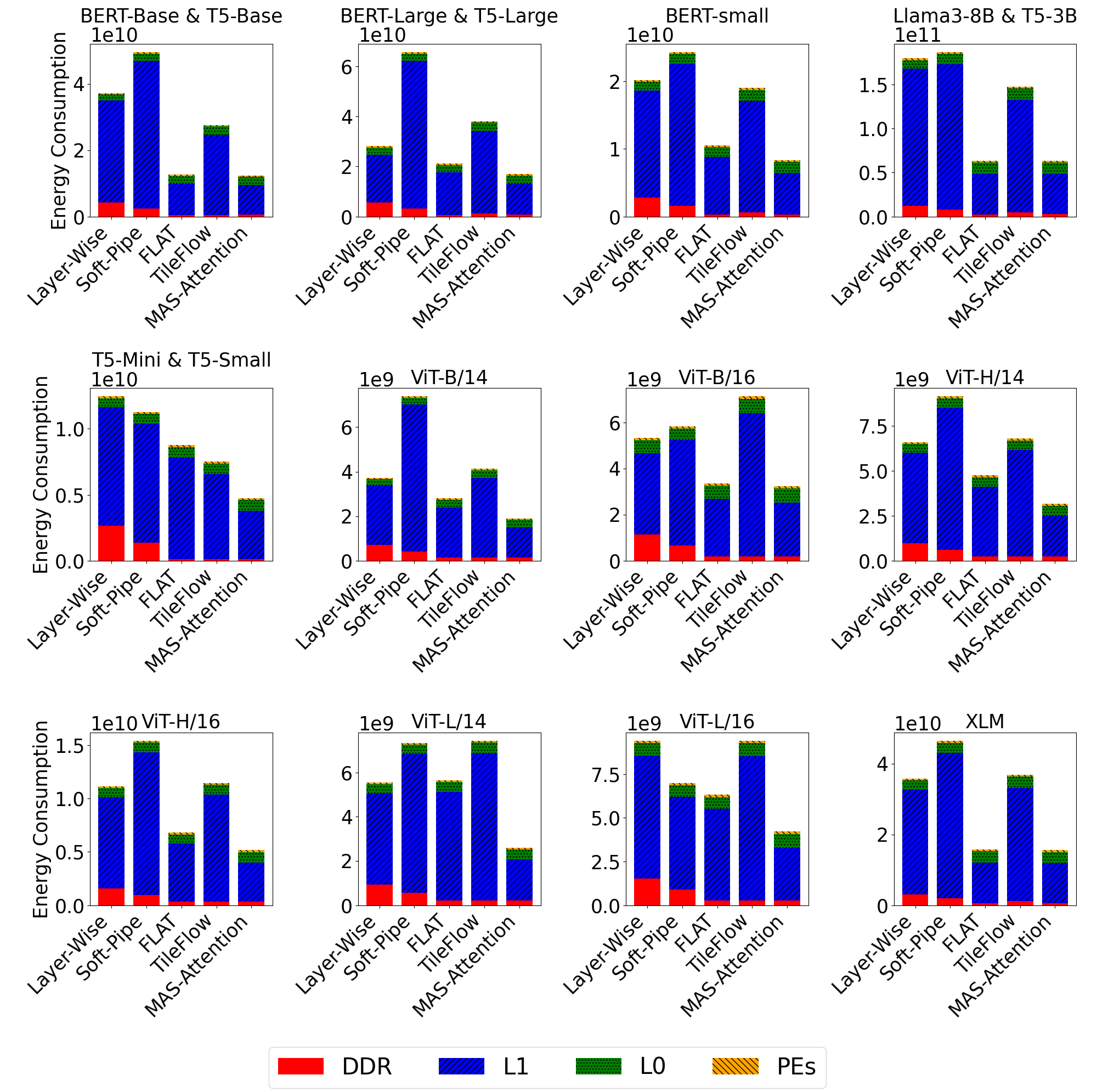}
  \vspace{-1.5em}
  \caption{Energy Consumption Breakdown for DDR, L1, L0 memories and PEs within MAC and VEC units Across Networks using Different Methods}
  \label{fig:energyConsumptionBreakdown_v2}
\end{figure}

In addition, we provide an energy consumption breakdown for each network on all algorithms as shown in \autoref{fig:energyConsumptionBreakdown_v2}, focusing on Off-Chip (DRAM) and On-Chip (L1, L0) memories, and PEs in MAC and Vector units. 

\subsubsection{Off-Chip Memory Energy Consumption}
Compared to Layer-Wise and Soft-Pipe methods, MAS-Attention significantly reduces off-chip energy consumption by minimizing DRAM accesses and eliminating the need to store intermediate $C$ and $P$ matrices off-chip. However, MAS-Attention's off-chip energy consumption in some cases is slightly higher than FLAT due to the need of reloading $K$ and $V$ matrices in the case of them being overwritten by the selective overwriting mechanism during pipelining. Soft-Pipe consumes more energy than MAS-Attention as it stores the $P$ matrix back to DRAM, but less than Layer-Wise as it does not store the $C$ matrix to DRAM.

\subsubsection{On-Chip Memory Energy Consumption}
Layer-Wise, Soft-Pipe and TileFlow usually consumes much more on-chip energy compared to MAS-Attention, indicating less efficient on-chip memory utilization. FLAT also show higher energy consumption than MAS-Attention but generally lower than Layer-Wise, Soft-Pipe and TileFlow. 

\subsubsection{PEs Energy Consumption}
Energy consumption in PEs remains constant across different algorithms for each network, as the actual computation required by different algorithms is the same, with differences only in the scheduling process.

\subsection{DRAM Access Analysis}

Since the FLAT method is most comparable to MAS-Attention in terms of both cycle and energy performance, we will focus on comparing the DRAM access between these two algorithms.

\subsubsection{DRAM Write Operations}
Both MAS-Attention and FLAT algorithms exhibit an identical number of write operations to DRAM. This uniformity arises because both algorithms confine their DRAM write operations to the final result of the attention block (\begin{math}O\end{math}), eschewing the need to write intermediate results to DRAM. Instead, these intermediate results are processed entirely on-chip, thereby minimizing off-chip memory accesses and enhancing overall efficiency.

\subsubsection{DRAM Read Operations}
Across the tested workloads, MAS-Attention matches FLAT in DRAM read operations but surpasses it for specific networks. Notably, for BERT-Base \& T5-Base ($1.5\times$), BERT-Large \& T5-Large ($1.5\times$), and Llama3-8B \& T5-3B ($1.49\times$), MAS-Attention shows increased DRAM read operations. This phenomenon arises because MAS-Attention requires reloading specific data chunks, particularly $K$ and $V$ matrices, which may have been overwritten during the pipelining stages on-chip. These matrices are reloaded from DRAM to resume the halted MAC operations, allowing the attention mechanism to maintain data dependencies and continue processing seamlessly. While this incurs additional DRAM reads, the proactive buffer overwriting mechanism maintains efficient on-chip memory usage and pipelined execution integrity, with total cycle counts and energy consumption still outperforming all other baselines.


\subsection{Impact of Search Algorithms on Tiling Optimization}

\begin{figure}[t]
  \centering
  \includegraphics[width=\columnwidth]{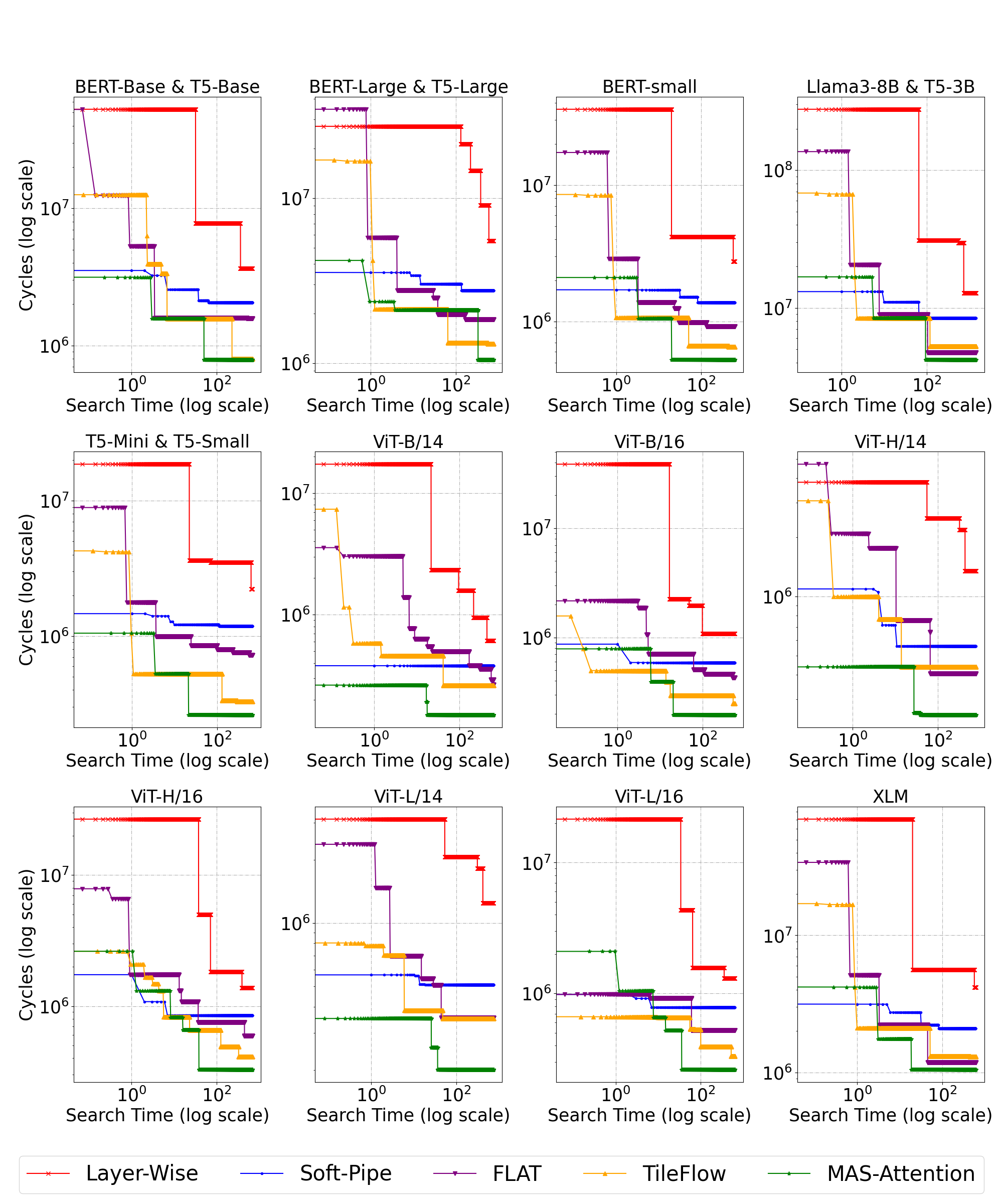}
  \vspace{-1.5em}
  \caption{Execution cycles vs. search time (both log scale) for different attention acceleration methods, demonstrating the impact of Genetic Algorithm (GA) and Monte Carlo Tree Search (MCTS) on each algorithm's efficiency}
  \label{fig:searchCycleTimeAnalysis}
\end{figure}

{Figure \ref{fig:searchCycleTimeAnalysis} illustrates the impact of employing MCTS and GA search algorithms in optimizing tile configurations for attention workloads. For clarity, the plot proportionally reduces the number of plotted lines to approximately 2K. It becomes evident that after around 10K iterations, each algorithm consistently converges toward optimal tiling parameters. Detailed final cycle counts and corresponding energy consumption metrics upon completion of the search are comprehensively listed in Tables \ref{tab:cycles_speedup} and \ref{tab:energy_savings}.} 
{FuseMax results in these tables and its original work were obtained via its manually selected tiling sizes for tensors on different memory levels, thus excluded from Figure \ref{fig:searchCycleTimeAnalysis} on search convergence.}

{To further underscore the efficacy of the proposed search scheme with MAS-Attention, notable cycle improvements include a 64.5× reduction for BERT-Base and T5-Base (from 50.33M to 0.78M), a 16.1× reduction for BERT-Large and T5-Large (from 16.77M to 1.04M), and a similar 16.1× improvement for BERT-Small (from 8.38M to 0.52M) as well as T5-Mini and T5-Small (from 4.19M to 0.26M). Furthermore, Vision transformer workloads demonstrate significant benefits with up to 66.2× speedup—ViT-{B,L,H}/14 see 49.7×/24.5×/24.6× (from 7.45M/4.91M/6.14M to 0.15M/0.20M/0.25M), ViT-{B,L,H}/16 show 66.2×/32.2×/32.8× (from 12.58M/8.38M/10.48M to 0.19M/0.26M/0.32M). Lastly, XLM sees a 32.2× drop (from 33.55M to 1.04M), further validating the broad applicability and robustness of the search-based optimization approach.}

\subsection{Limitations}
    On the simulated edge hardware, MAS-Attention can handle a maximum sequence length of approximately 1 million tokens in half precision (FP16), which is half the maximum sequence length that FLAT can handle. The computation of $P_i$ happens in parallel with either $O_{i-1}$ or $C_{i+1}$. Since Softmax operates row-wise, at least one row is used in the computation of $P_i$. In the case of $P_i$ computed in parallel with $O_{i-1} = P_{i-1}V$, $O_{i-1}$ requires at least one entire row of $P_{i-1}$ to be calculated. Also, in the case of $P_i$ computed in parallel with $C_{i+1}=Q_{i+1}K^T$, one entire row of $C_{i+1}$ is computed and written on-chip. In both scenarios, on-chip memory should have the capacity for either $P_i$ and $P_{i-1}$ or $P_i$ and $C_{i+1}$. In the case of half precision with a sequence length of 1M, one row of $P_i$, $P_{i-1}$, and $C_{i+1}$ consumes 2MB each on-chip, which fits within the 5MB on-chip cache size in either scenario. Since FLAT does not employ such a pipelining scheme and operates sequentially, it can handle a sequence length of 2 million tokens. In this condition, one row of $P_i$ consumes 4MB on-chip, which can be managed by the 5MB on-chip cache size in the simulated edge device.



    {Furthermore, MAS-Attention's stream processing efficiency relies on the availability of separate compute engines for MatMul and Softmax operations, leveraging dedicated MAC and VEC units for parallel execution. Therefore, MAS-Attention remains particularly effective on architectures with distinct heterogeneous compute resources—a design choice becoming increasingly common in modern edge accelerators to optimize for energy efficiency.}

%% file: sections/conclusion.tex
\label{sec:conclusion}
In this paper, we propose MAS-Attention dataflow to accelerate attention mechanism on resource-constrained edge devices. Our approach uses a stream processing scheme to execute tiled MatMul and Softmax workloads in a pipelined manner, with MAC and VEC units operating in parallel. A multi-tiered tiling strategy ensures balanced workloads for efficient pipelined attention execution. Additionally, our proactive buffer overwrite strategy enhances on-chip memory utilization by freeing up buffer space when it runs out of memory, such as with longer input sequences. 
While this strategy increases off-chip memory reads, MAS-Attention achieves superior speedup and energy savings over previous methods like Layer-wise, Soft-Pipe, FLAT, and TileFlow, on both simulated and real edge devices.

Future work will extend MAS-Attention to support training, which adds complexity in backpropagation that challenges efficient workload management on resource-constrained edge devices.


%% file: main.bbl
\begin{thebibliography}{79}
\providecommand{\natexlab}[1]{#1}
\providecommand{\url}[1]{\texttt{#1}}
\expandafter\ifx\csname urlstyle\endcsname\relax
  \providecommand{\doi}[1]{doi: #1}\else
  \providecommand{\doi}{doi: \begingroup \urlstyle{rm}\Url}\fi

\bibitem[Nvi()]{NvidiaTensorRT}
{Nvidia, TensorRT}.
\newblock \url{https://docs.nvidia.com/deeplearning/tensorrt/archives/tensorrt-803/best-practices/index.html}.
\newblock Accessed: 2024.

\bibitem[app({\natexlab{a}})]{appleAccelerateF}
{Apple, Accelerate Framework}.
\newblock \url{https://developer.apple.com/documentation/accelerate/bnns}, {\natexlab{a}}.
\newblock Accessed: 2023.

\bibitem[app({\natexlab{b}})]{appleCoreML}
{Apple, Core ML Tools}.
\newblock \url{https://apple.github.io/coremltools/docs-guides/source/opt-palettization-overview.html}, {\natexlab{b}}.
\newblock Accessed: 2023.

\bibitem[app({\natexlab{c}})]{appleMetalPSG}
{Apple, Metal Performance Shaders Graph}.
\newblock \url{https://developer.apple.com/documentation/metalperformanceshadersgraph}, {\natexlab{c}}.
\newblock Accessed: 2023.

\bibitem[dao()]{dao2023flashdecoding}
{T. Dao, D. Haziza, F. Massa, G. Sizov, Flash-Decoding for long-context inference}.
\newblock \url{https://crfm.stanford.edu/2023/10/12/flashdecoding.html}.
\newblock Accessed: 2023-10-12.

\bibitem[goo()]{googleXLA}
{Google, TensorFlow XLA}.
\newblock \url{https://www.tensorflow}.
\newblock Accessed: 2021.

\bibitem[Aminabadi et~al.(2022)Aminabadi, Rajbhandari, Awan, Li, Li, Zheng, Ruwase, Smith, Zhang, Rasley, et~al.]{aminabadi2022deepspeed}
Aminabadi, R.~Y., Rajbhandari, S., Awan, A.~A., Li, C., Li, D., Zheng, E., Ruwase, O., Smith, S., Zhang, M., Rasley, J., et~al.
\newblock Deepspeed-inference: enabling efficient inference of transformer models at unprecedented scale.
\newblock In \emph{SC22: International Conference for High Performance Computing, Networking, Storage and Analysis}, pp.\  1--15. IEEE, 2022.

\bibitem[Baghdadi et~al.(2019)Baghdadi, Ray, Romdhane, Del~Sozzo, Akkas, Zhang, Suriana, Kamil, and Amarasinghe]{baghdadi2019tiramisu}
Baghdadi, R., Ray, J., Romdhane, M.~B., Del~Sozzo, E., Akkas, A., Zhang, Y., Suriana, P., Kamil, S., and Amarasinghe, S.
\newblock Tiramisu: A polyhedral compiler for expressing fast and portable code.
\newblock In \emph{2019 IEEE/ACM International Symposium on Code Generation and Optimization (CGO)}, pp.\  193--205. IEEE, 2019.

\bibitem[Chen et~al.(2018{\natexlab{a}})Chen, Moreau, Jiang, Zheng, Yan, Shen, Cowan, Wang, Hu, Ceze, et~al.]{chen2018tvm}
Chen, T., Moreau, T., Jiang, Z., Zheng, L., Yan, E., Shen, H., Cowan, M., Wang, L., Hu, Y., Ceze, L., et~al.
\newblock $\{$TVM$\}$: An automated $\{$End-to-End$\}$ optimizing compiler for deep learning.
\newblock In \emph{13th USENIX Symposium on Operating Systems Design and Implementation (OSDI 18)}, pp.\  578--594, 2018{\natexlab{a}}.

\bibitem[Chen et~al.(2018{\natexlab{b}})Chen, Zheng, Yan, Jiang, Moreau, Ceze, Guestrin, and Krishnamurthy]{chen2018learning}
Chen, T., Zheng, L., Yan, E., Jiang, Z., Moreau, T., Ceze, L., Guestrin, C., and Krishnamurthy, A.
\newblock Learning to optimize tensor programs.
\newblock \emph{Advances in Neural Information Processing Systems}, 31, 2018{\natexlab{b}}.

\bibitem[Chen et~al.(2019)Chen, Yang, Emer, and Sze]{chen2019eyeriss}
Chen, Y.-H., Yang, T.-J., Emer, J., and Sze, V.
\newblock Eyeriss v2: A flexible accelerator for emerging deep neural networks on mobile devices.
\newblock \emph{IEEE Journal on Emerging and Selected Topics in Circuits and Systems}, 9\penalty0 (2):\penalty0 292--308, 2019.

\bibitem[Cho et~al.(2021)Cho, Vahid, Adya, and Rastegari]{cho2021dkm}
Cho, M., Vahid, K.~A., Adya, S., and Rastegari, M.
\newblock Dkm: Differentiable k-means clustering layer for neural network compression.
\newblock \emph{arXiv preprint arXiv:2108.12659}, 2021.

\bibitem[Cho et~al.(2024)Cho, Rastegari, and Naik]{cho2024kv}
Cho, M., Rastegari, M., and Naik, D.
\newblock Kv-runahead: Scalable causal llm inference by parallel key-value cache generation.
\newblock \emph{arXiv preprint arXiv:2405.05329}, 2024.

\bibitem[Dao(2023)]{dao2023flashattention}
Dao, T.
\newblock Flashattention-2: Faster attention with better parallelism and work partitioning.
\newblock \emph{arXiv preprint arXiv:2307.08691}, 2023.

\bibitem[Dao et~al.(2022)Dao, Fu, Ermon, Rudra, and R{\'e}]{dao2022flashattention}
Dao, T., Fu, D., Ermon, S., Rudra, A., and R{\'e}, C.
\newblock Flashattention: Fast and memory-efficient exact attention with io-awareness.
\newblock \emph{Advances in Neural Information Processing Systems}, 35:\penalty0 16344--16359, 2022.

\bibitem[Devlin et~al.(2018)Devlin, Chang, Lee, and Toutanova]{devlin2018bert}
Devlin, J., Chang, M.-W., Lee, K., and Toutanova, K.
\newblock Bert: Pre-training of deep bidirectional transformers for language understanding.
\newblock \emph{arXiv preprint arXiv:1810.04805}, 2018.

\bibitem[Dosovitskiy et~al.(2020)Dosovitskiy, Beyer, Kolesnikov, Weissenborn, Zhai, Unterthiner, Dehghani, Minderer, Heigold, Gelly, et~al.]{dosovitskiy2020image}
Dosovitskiy, A., Beyer, L., Kolesnikov, A., Weissenborn, D., Zhai, X., Unterthiner, T., Dehghani, M., Minderer, M., Heigold, G., Gelly, S., et~al.
\newblock An image is worth 16x16 words: Transformers for image recognition at scale.
\newblock \emph{arXiv preprint arXiv:2010.11929}, 2020.

\bibitem[Esser et~al.(2024)Esser, Kulal, Blattmann, Entezari, M{\"u}ller, Saini, Levi, Lorenz, Sauer, Boesel, et~al.]{esser2024scaling}
Esser, P., Kulal, S., Blattmann, A., Entezari, R., M{\"u}ller, J., Saini, H., Levi, Y., Lorenz, D., Sauer, A., Boesel, F., et~al.
\newblock Scaling rectified flow transformers for high-resolution image synthesis.
\newblock In \emph{Forty-first International Conference on Machine Learning}, 2024.

\bibitem[Fan et~al.(2025)Fan, Ma, Wu, Du, Li, Gao, and Li]{fan2025videoagent}
Fan, Y., Ma, X., Wu, R., Du, Y., Li, J., Gao, Z., and Li, Q.
\newblock Videoagent: A memory-augmented multimodal agent for video understanding.
\newblock In \emph{European Conference on Computer Vision}, pp.\  75--92. Springer, 2025.

\bibitem[Ganesh et~al.(2021)Ganesh, Chen, Lou, Khan, Yang, Sajjad, Nakov, Chen, and Winslett]{ganesh2021compressing}
Ganesh, P., Chen, Y., Lou, X., Khan, M.~A., Yang, Y., Sajjad, H., Nakov, P., Chen, D., and Winslett, M.
\newblock Compressing large-scale transformer-based models: A case study on bert.
\newblock \emph{Transactions of the Association for Computational Linguistics}, 9:\penalty0 1061--1080, 2021.

\bibitem[Gao et~al.(1993)Gao, Olsen, Sarkar, and Thekkath]{gao1993collective}
Gao, G., Olsen, R., Sarkar, V., and Thekkath, R.
\newblock Collective loop fusion for array contraction.
\newblock In \emph{Languages and Compilers for Parallel Computing: 5th International Workshop New Haven, Connecticut, USA, August 3--5, 1992 Proceedings 5}, pp.\  281--295. Springer, 1993.

\bibitem[Glaese et~al.(2022)Glaese, McAleese, Tr{e}bacz, Aslanides, Firoiu, Ewalds, Rauh, Weidinger, Chadwick, Thacker, et~al.]{glaese2022improving}
Glaese, A., McAleese, N., Tr{e}bacz, M., Aslanides, J., Firoiu, V., Ewalds, T., Rauh, M., Weidinger, L., Chadwick, M., Thacker, P., et~al.
\newblock Improving alignment of dialogue agents via targeted human judgements.
\newblock \emph{arXiv preprint arXiv:2209.14375}, 2022.

\bibitem[Gupta et~al.(2024)Gupta, Jaddipal, Prabhala, Paul, and Von~Platen]{gupta2024progressive}
Gupta, Y., Jaddipal, V.~V., Prabhala, H., Paul, S., and Von~Platen, P.
\newblock Progressive knowledge distillation of stable diffusion xl using layer level loss.
\newblock \emph{arXiv preprint arXiv:2401.02677}, 2024.

\bibitem[Hong et~al.(2023)Hong, Dai, Xu, Mao, Li, Liu, Chen, Dong, and Wang]{hong2023flashdecoding++}
Hong, K., Dai, G., Xu, J., Mao, Q., Li, X., Liu, J., Chen, K., Dong, H., and Wang, Y.
\newblock Flashdecoding++: Faster large language model inference on gpus.
\newblock \emph{arXiv preprint arXiv:2311.01282}, 2023.

\bibitem[Huang et~al.(2024)Huang, Zhang, Zheng, You, Wang, Qian, and Xu]{huang2024knowledge}
Huang, T., Zhang, Y., Zheng, M., You, S., Wang, F., Qian, C., and Xu, C.
\newblock Knowledge diffusion for distillation.
\newblock \emph{Advances in Neural Information Processing Systems}, 36, 2024.

\bibitem[Ivanov et~al.(2021)Ivanov, Dryden, Ben-Nun, Li, and Hoefler]{ivanov2021data}
Ivanov, A., Dryden, N., Ben-Nun, T., Li, S., and Hoefler, T.
\newblock Data movement is all you need: A case study on optimizing transformers.
\newblock \emph{Proceedings of Machine Learning and Systems}, 3:\penalty0 711--732, 2021.

\bibitem[Jouppi et~al.(2017)Jouppi, Young, Patil, Patterson, Agrawal, Bajwa, Bates, Bhatia, Boden, Borchers, et~al.]{jouppi2017datacenter}
Jouppi, N.~P., Young, C., Patil, N., Patterson, D., Agrawal, G., Bajwa, R., Bates, S., Bhatia, S., Boden, N., Borchers, A., et~al.
\newblock In-datacenter performance analysis of a tensor processing unit.
\newblock In \emph{Proceedings of the 44th annual international symposium on computer architecture}, pp.\  1--12, 2017.

\bibitem[Jouppi et~al.(2020)Jouppi, Yoon, Kurian, Li, Patil, Laudon, Young, and Patterson]{jouppi2020domain}
Jouppi, N.~P., Yoon, D.~H., Kurian, G., Li, S., Patil, N., Laudon, J., Young, C., and Patterson, D.
\newblock A domain-specific supercomputer for training deep neural networks.
\newblock \emph{Communications of the ACM}, 63\penalty0 (7):\penalty0 67--78, 2020.

\bibitem[Kao et~al.(2023)Kao, Subramanian, Agrawal, Yazdanbakhsh, and Krishna]{kao2023flat}
Kao, S.-C., Subramanian, S., Agrawal, G., Yazdanbakhsh, A., and Krishna, T.
\newblock Flat: An optimized dataflow for mitigating attention bottlenecks.
\newblock In \emph{Proceedings of the 28th ACM International Conference on Architectural Support for Programming Languages and Operating Systems, Volume 2}, pp.\  295--310, 2023.

\bibitem[Kaplan et~al.(2020)Kaplan, McCandlish, Henighan, Brown, Chess, Child, Gray, Radford, Wu, and Amodei]{kaplan2020scaling}
Kaplan, J., McCandlish, S., Henighan, T., Brown, T.~B., Chess, B., Child, R., Gray, S., Radford, A., Wu, J., and Amodei, D.
\newblock Scaling laws for neural language models.
\newblock \emph{arXiv preprint arXiv:2001.08361}, 2020.

\bibitem[Kirk et~al.(2007)]{kirk2007nvidia}
Kirk, D. et~al.
\newblock Nvidia cuda software and gpu parallel computing architecture.
\newblock In \emph{ISMM}, volume~7, pp.\  103--104, 2007.

\bibitem[Kitaev et~al.(2020)Kitaev, Kaiser, and Levskaya]{kitaev2020reformer}
Kitaev, N., Kaiser, {\L}., and Levskaya, A.
\newblock Reformer: The efficient transformer.
\newblock \emph{arXiv preprint arXiv:2001.04451}, 2020.

\bibitem[Kjolstad et~al.(2017)Kjolstad, Kamil, Chou, Lugato, and Amarasinghe]{kjolstad2017tensor}
Kjolstad, F., Kamil, S., Chou, S., Lugato, D., and Amarasinghe, S.
\newblock The tensor algebra compiler.
\newblock \emph{Proceedings of the ACM on Programming Languages}, 1\penalty0 (OOPSLA):\penalty0 1--29, 2017.

\bibitem[Kwon et~al.(2018)Kwon, Samajdar, and Krishna]{kwon2018maeri}
Kwon, H., Samajdar, A., and Krishna, T.
\newblock Maeri: Enabling flexible dataflow mapping over dnn accelerators via reconfigurable interconnects.
\newblock \emph{ACM SIGPLAN Notices}, 53\penalty0 (2):\penalty0 461--475, 2018.

\bibitem[Kwon et~al.(2023)Kwon, Li, Zhuang, Sheng, Zheng, Yu, Gonzalez, Zhang, and Stoica]{kwon2023efficient}
Kwon, W., Li, Z., Zhuang, S., Sheng, Y., Zheng, L., Yu, C.~H., Gonzalez, J., Zhang, H., and Stoica, I.
\newblock Efficient memory management for large language model serving with pagedattention.
\newblock In \emph{Proceedings of the 29th Symposium on Operating Systems Principles}, pp.\  611--626, 2023.

\bibitem[Lample \& Conneau(2019)Lample and Conneau]{lample2019cross}
Lample, G. and Conneau, A.
\newblock Cross-lingual language model pretraining.
\newblock \emph{arXiv preprint arXiv:1901.07291}, 2019.

\bibitem[Li et~al.(2024{\natexlab{a}})Li, Gan, Yang, Yang, Li, Wang, Gao, et~al.]{li2024multimodal}
Li, C., Gan, Z., Yang, Z., Yang, J., Li, L., Wang, L., Gao, J., et~al.
\newblock Multimodal foundation models: From specialists to general-purpose assistants.
\newblock \emph{Foundations and Trends{\textregistered} in Computer Graphics and Vision}, 16\penalty0 (1-2):\penalty0 1--214, 2024{\natexlab{a}}.

\bibitem[Li et~al.(2024{\natexlab{b}})Li, Qin, Mei, Cui, Song, Chen, Zhang, Du, Cheng, Jin, et~al.]{li2024onednn}
Li, J., Qin, Z., Mei, Y., Cui, J., Song, Y., Chen, C., Zhang, Y., Du, L., Cheng, X., Jin, B., et~al.
\newblock onednn graph compiler: A hybrid approach for high-performance deep learning compilation.
\newblock In \emph{2024 IEEE/ACM International Symposium on Code Generation and Optimization (CGO)}, pp.\  460--470. IEEE, 2024{\natexlab{b}}.

\bibitem[Li et~al.(2022)Li, Xu, Zhang, Cao, Gao, and Guo]{li2022q}
Li, Y., Xu, S., Zhang, B., Cao, X., Gao, P., and Guo, G.
\newblock Q-vit: Accurate and fully quantized low-bit vision transformer.
\newblock \emph{Advances in neural information processing systems}, 35:\penalty0 34451--34463, 2022.

\bibitem[Li \& Gu(2023)Li and Gu]{li2023vit}
Li, Z. and Gu, Q.
\newblock I-vit: Integer-only quantization for efficient vision transformer inference.
\newblock In \emph{Proceedings of the IEEE/CVF International Conference on Computer Vision}, pp.\  17065--17075, 2023.

\bibitem[Liao et~al.(2019)Liao, Tu, Xia, and Zhou]{liao2019davinci}
Liao, H., Tu, J., Xia, J., and Zhou, X.
\newblock Davinci: A scalable architecture for neural network computing.
\newblock In \emph{2019 IEEE Hot Chips 31 Symposium (HCS)}, pp.\  1--44. IEEE Computer Society, 2019.

\bibitem[Lin et~al.(2021)Lin, Zhang, Sun, Li, and Zhou]{lin2021fq}
Lin, Y., Zhang, T., Sun, P., Li, Z., and Zhou, S.
\newblock Fq-vit: Post-training quantization for fully quantized vision transformer.
\newblock \emph{arXiv preprint arXiv:2111.13824}, 2021.

\bibitem[Liu et~al.(2023)Liu, Zaharia, and Abbeel]{liu2023ring}
Liu, H., Zaharia, M., and Abbeel, P.
\newblock Ring attention with blockwise transformers for near-infinite context.
\newblock \emph{arXiv preprint arXiv:2310.01889}, 2023.

\bibitem[Liu et~al.(2021)Liu, Wang, Han, Zhang, Ma, and Gao]{liu2021post}
Liu, Z., Wang, Y., Han, K., Zhang, W., Ma, S., and Gao, W.
\newblock Post-training quantization for vision transformer.
\newblock \emph{Advances in Neural Information Processing Systems}, 34:\penalty0 28092--28103, 2021.

\bibitem[Mao et~al.(2021)Mao, Yang, Li, Li, and Chen]{mao2021tprune}
Mao, J., Yang, H., Li, A., Li, H., and Chen, Y.
\newblock Tprune: Efficient transformer pruning for mobile devices.
\newblock \emph{ACM Transactions on Cyber-Physical Systems}, 5\penalty0 (3):\penalty0 1--22, 2021.

\bibitem[Mehta et~al.(2024)Mehta, Sekhavat, Cao, Horton, Jin, Sun, Mirzadeh, Najibi, Belenko, Zatloukal, et~al.]{mehta2024openelm}
Mehta, S., Sekhavat, M.~H., Cao, Q., Horton, M., Jin, Y., Sun, C., Mirzadeh, I., Najibi, M., Belenko, D., Zatloukal, P., et~al.
\newblock Openelm: An efficient language model family with open-source training and inference framework.
\newblock \emph{arXiv preprint arXiv:2404.14619}, 2024.

\bibitem[Mei et~al.(2023)Mei, Goetschalckx, Symons, and Verhelst]{mei2023defines}
Mei, L., Goetschalckx, K., Symons, A., and Verhelst, M.
\newblock Defines: Enabling fast exploration of the depth-first scheduling space for dnn accelerators through analytical modeling.
\newblock In \emph{2023 IEEE International Symposium on High-Performance Computer Architecture (HPCA)}, pp.\  570--583. IEEE, 2023.

\bibitem[Narayanan et~al.(2021)Narayanan, Shoeybi, Casper, LeGresley, Patwary, Korthikanti, Vainbrand, Kashinkunti, Bernauer, Catanzaro, et~al.]{narayanan2021efficient}
Narayanan, D., Shoeybi, M., Casper, J., LeGresley, P., Patwary, M., Korthikanti, V., Vainbrand, D., Kashinkunti, P., Bernauer, J., Catanzaro, B., et~al.
\newblock Efficient large-scale language model training on gpu clusters using megatron-lm.
\newblock In \emph{Proceedings of the International Conference for High Performance Computing, Networking, Storage and Analysis}, pp.\  1--15, 2021.

\bibitem[Nayak et~al.(2024)Nayak, Wu, Odemuyiwa, Pellauer, Emer, and Fletcher]{nayak2024fusemax}
Nayak, N., Wu, X., Odemuyiwa, T.~O., Pellauer, M., Emer, J.~S., and Fletcher, C.~W.
\newblock Fusemax: Leveraging extended einsums to optimize attention accelerator design.
\newblock \emph{arXiv preprint arXiv:2406.10491}, 2024.

\bibitem[Niu et~al.(2021)Niu, Guan, Wang, Agrawal, and Ren]{niu2021dnnfusion}
Niu, W., Guan, J., Wang, Y., Agrawal, G., and Ren, B.
\newblock Dnnfusion: accelerating deep neural networks execution with advanced operator fusion.
\newblock In \emph{Proceedings of the 42nd ACM SIGPLAN International Conference on Programming Language Design and Implementation}, pp.\  883--898, 2021.

\bibitem[Ouyang et~al.(2022)Ouyang, Wu, Jiang, Almeida, Wainwright, Mishkin, Zhang, Agarwal, Slama, Ray, et~al.]{ouyang2022training}
Ouyang, L., Wu, J., Jiang, X., Almeida, D., Wainwright, C., Mishkin, P., Zhang, C., Agarwal, S., Slama, K., Ray, A., et~al.
\newblock Training language models to follow instructions with human feedback.
\newblock \emph{Advances in neural information processing systems}, 35:\penalty0 27730--27744, 2022.

\bibitem[Parashar et~al.(2019)Parashar, Raina, Shao, Chen, Ying, Mukkara, Venkatesan, Khailany, Keckler, and Emer]{parashar2019timeloop}
Parashar, A., Raina, P., Shao, Y.~S., Chen, Y.-H., Ying, V.~A., Mukkara, A., Venkatesan, R., Khailany, B., Keckler, S.~W., and Emer, J.
\newblock Timeloop: A systematic approach to dnn accelerator evaluation.
\newblock In \emph{2019 IEEE international symposium on performance analysis of systems and software (ISPASS)}, pp.\  304--315. IEEE, 2019.

\bibitem[Patel et~al.(2023)Patel, Choukse, Zhang, Shah, Goiri, Maleki, and Bianchini]{patel2023splitwise}
Patel, P., Choukse, E., Zhang, C., Shah, A., Goiri, {\'I}., Maleki, S., and Bianchini, R.
\newblock Splitwise: Efficient generative llm inference using phase splitting.
\newblock \emph{Power}, 400\penalty0 (700W):\penalty0 1--75, 2023.

\bibitem[Peebles \& Xie(2023)Peebles and Xie]{peebles2023scalable}
Peebles, W. and Xie, S.
\newblock Scalable diffusion models with transformers.
\newblock In \emph{Proceedings of the IEEE/CVF International Conference on Computer Vision}, pp.\  4195--4205, 2023.

\bibitem[Peng et~al.(2021)Peng, Huang, Geng, Li, Jiang, Liu, Wang, and Ding]{peng2021accelerating}
Peng, H., Huang, S., Geng, T., Li, A., Jiang, W., Liu, H., Wang, S., and Ding, C.
\newblock Accelerating transformer-based deep learning models on fpgas using column balanced block pruning.
\newblock In \emph{2021 22nd International Symposium on Quality Electronic Design (ISQED)}, pp.\  142--148. IEEE, 2021.

\bibitem[Piao et~al.(2022)Piao, Cho, and Kang]{piao2022sensimix}
Piao, T., Cho, I., and Kang, U.
\newblock Sensimix: Sensitivity-aware 8-bit index \& 1-bit value mixed precision quantization for bert compression.
\newblock \emph{PloS one}, 17\penalty0 (4):\penalty0 e0265621, 2022.

\bibitem[Poole et~al.(2022)Poole, Jain, Barron, and Mildenhall]{poole2022dreamfusion}
Poole, B., Jain, A., Barron, J.~T., and Mildenhall, B.
\newblock Dreamfusion: Text-to-3d using 2d diffusion.
\newblock \emph{arXiv preprint arXiv:2209.14988}, 2022.

\bibitem[Radford et~al.(2018)Radford, Narasimhan, Salimans, Sutskever, et~al.]{radford2018improving}
Radford, A., Narasimhan, K., Salimans, T., Sutskever, I., et~al.
\newblock Improving language understanding by generative pre-training.
\newblock 2018.

\bibitem[Raffel et~al.(2020)Raffel, Shazeer, Roberts, Lee, Narang, Matena, Zhou, Li, and Liu]{raffel2020exploring}
Raffel, C., Shazeer, N., Roberts, A., Lee, K., Narang, S., Matena, M., Zhou, Y., Li, W., and Liu, P.~J.
\newblock Exploring the limits of transfer learning with a unified text-to-text transformer.
\newblock \emph{Journal of machine learning research}, 21\penalty0 (140):\penalty0 1--67, 2020.

\bibitem[Rasley et~al.(2020)Rasley, Rajbhandari, Ruwase, and He]{rasley2020deepspeed}
Rasley, J., Rajbhandari, S., Ruwase, O., and He, Y.
\newblock Deepspeed: System optimizations enable training deep learning models with over 100 billion parameters.
\newblock In \emph{Proceedings of the 26th ACM SIGKDD International Conference on Knowledge Discovery \& Data Mining}, pp.\  3505--3506, 2020.

\bibitem[Shah et~al.(2024)Shah, Bikshandi, Zhang, Thakkar, Ramani, and Dao]{shah2024flashattention}
Shah, J., Bikshandi, G., Zhang, Y., Thakkar, V., Ramani, P., and Dao, T.
\newblock Flashattention-3: Fast and accurate attention with asynchrony and low-precision.
\newblock \emph{arXiv preprint arXiv:2407.08608}, 2024.

\bibitem[Shoeybi et~al.(2019)Shoeybi, Patwary, Puri, LeGresley, Casper, and Catanzaro]{shoeybi2019megatron}
Shoeybi, M., Patwary, M., Puri, R., LeGresley, P., Casper, J., and Catanzaro, B.
\newblock Megatron-lm: Training multi-billion parameter language models using model parallelism.
\newblock \emph{arXiv preprint arXiv:1909.08053}, 2019.

\bibitem[Sun et~al.(2019)Sun, Cheng, Gan, and Liu]{sun2019patient}
Sun, S., Cheng, Y., Gan, Z., and Liu, J.
\newblock Patient knowledge distillation for bert model compression.
\newblock \emph{arXiv preprint arXiv:1908.09355}, 2019.

\bibitem[Tabani et~al.(2021)Tabani, Balasubramaniam, Marzban, Arani, and Zonooz]{tabani2021improving}
Tabani, H., Balasubramaniam, A., Marzban, S., Arani, E., and Zonooz, B.
\newblock Improving the efficiency of transformers for resource-constrained devices.
\newblock In \emph{2021 24th Euromicro Conference on Digital System Design (DSD)}, pp.\  449--456. IEEE, 2021.

\bibitem[Touvron et~al.(2023)Touvron, Lavril, Izacard, Martinet, Lachaux, Lacroix, Rozi{\`e}re, Goyal, Hambro, Azhar, et~al.]{touvron2023llama}
Touvron, H., Lavril, T., Izacard, G., Martinet, X., Lachaux, M.-A., Lacroix, T., Rozi{\`e}re, B., Goyal, N., Hambro, E., Azhar, F., et~al.
\newblock Llama: Open and efficient foundation language models.
\newblock \emph{arXiv preprint arXiv:2302.13971}, 2023.

\bibitem[Vaswani et~al.(2017)Vaswani, Shazeer, Parmar, Uszkoreit, Jones, Gomez, Kaiser, and Polosukhin]{vaswani2017attention}
Vaswani, A., Shazeer, N., Parmar, N., Uszkoreit, J., Jones, L., Gomez, A.~N., Kaiser, {\L}., and Polosukhin, I.
\newblock Attention is all you need.
\newblock \emph{Advances in neural information processing systems}, 30, 2017.

\bibitem[Wang et~al.(2024)Wang, Xu, Ye, Yan, Shen, Zhang, Huang, and Sang]{wang2024mobile}
Wang, J., Xu, H., Ye, J., Yan, M., Shen, W., Zhang, J., Huang, F., and Sang, J.
\newblock Mobile-agent: Autonomous multi-modal mobile device agent with visual perception.
\newblock \emph{arXiv preprint arXiv:2401.16158}, 2024.

\bibitem[Wang et~al.(2022)Wang, Liu, Venkataramani, Sen, Chen, El~Maghraoui, Srinivasan, and Chang]{wang2022deep}
Wang, N., Liu, C.-C.~C., Venkataramani, S., Sen, S., Chen, C.-Y., El~Maghraoui, K., Srinivasan, V.~V., and Chang, L.
\newblock Deep compression of pre-trained transformer models.
\newblock \emph{Advances in Neural Information Processing Systems}, 35:\penalty0 14140--14154, 2022.

\bibitem[Wang et~al.(2020{\natexlab{a}})Wang, Zhou, Gan, Chen, Fang, Sun, Cheng, and Liu]{wang2020cluster}
Wang, S., Zhou, L., Gan, Z., Chen, Y.-C., Fang, Y., Sun, S., Cheng, Y., and Liu, J.
\newblock Cluster-former: Clustering-based sparse transformer for long-range dependency encoding.
\newblock \emph{arXiv preprint arXiv:2009.06097}, 2020{\natexlab{a}}.

\bibitem[Wang et~al.(2020{\natexlab{b}})Wang, Bao, Huang, Dong, and Wei]{wang2020minilmv2}
Wang, W., Bao, H., Huang, S., Dong, L., and Wei, F.
\newblock Minilmv2: Multi-head self-attention relation distillation for compressing pretrained transformers.
\newblock \emph{arXiv preprint arXiv:2012.15828}, 2020{\natexlab{b}}.

\bibitem[Wang et~al.(2020{\natexlab{c}})Wang, Wei, Dong, Bao, Yang, and Zhou]{wang2020minilm}
Wang, W., Wei, F., Dong, L., Bao, H., Yang, N., and Zhou, M.
\newblock Minilm: Deep self-attention distillation for task-agnostic compression of pre-trained transformers.
\newblock \emph{Advances in Neural Information Processing Systems}, 33:\penalty0 5776--5788, 2020{\natexlab{c}}.

\bibitem[Wu et~al.(2019)Wu, Emer, and Sze]{wu2019accelergy}
Wu, Y.~N., Emer, J.~S., and Sze, V.
\newblock Accelergy: An architecture-level energy estimation methodology for accelerator designs.
\newblock In \emph{2019 IEEE/ACM International Conference on Computer-Aided Design (ICCAD)}, pp.\  1--8. IEEE, 2019.

\bibitem[Yao et~al.(2022)Yao, Yazdani~Aminabadi, Zhang, Wu, Li, and He]{yao2022zeroquant}
Yao, Z., Yazdani~Aminabadi, R., Zhang, M., Wu, X., Li, C., and He, Y.
\newblock Zeroquant: Efficient and affordable post-training quantization for large-scale transformers.
\newblock \emph{Advances in Neural Information Processing Systems}, 35:\penalty0 27168--27183, 2022.

\bibitem[Yu et~al.(2023)Yu, Chen, Gan, and Fan]{yu2023boost}
Yu, C., Chen, T., Gan, Z., and Fan, J.
\newblock Boost vision transformer with gpu-friendly sparsity and quantization.
\newblock In \emph{Proceedings of the IEEE/CVF Conference on Computer Vision and Pattern Recognition}, pp.\  22658--22668, 2023.

\bibitem[Yu et~al.(2022{\natexlab{a}})Yu, Huang, Wang, Cheng, Chu, and Cui]{yu2022width}
Yu, F., Huang, K., Wang, M., Cheng, Y., Chu, W., and Cui, L.
\newblock Width \& depth pruning for vision transformers.
\newblock In \emph{Proceedings of the AAAI Conference on Artificial Intelligence}, volume~36, pp.\  3143--3151, 2022{\natexlab{a}}.

\bibitem[Yu et~al.(2022{\natexlab{b}})Yu, Chen, Shen, Yuan, Tan, Yang, Liu, and Wang]{yu2022unified}
Yu, S., Chen, T., Shen, J., Yuan, H., Tan, J., Yang, S., Liu, J., and Wang, Z.
\newblock Unified visual transformer compression.
\newblock \emph{arXiv preprint arXiv:2203.08243}, 2022{\natexlab{b}}.

\bibitem[Zhang et~al.(2023)Zhang, Yang, Liu, Han, Chen, Huang, Fu, and Yu]{zhang2023appagent}
Zhang, C., Yang, Z., Liu, J., Han, Y., Chen, X., Huang, Z., Fu, B., and Yu, G.
\newblock Appagent: Multimodal agents as smartphone users.
\newblock \emph{arXiv preprint arXiv:2312.13771}, 2023.

\bibitem[Zheng et~al.(2023)Zheng, Chen, Gao, Jia, Sun, Wang, and Liang]{zheng2023tileflow}
Zheng, S., Chen, S., Gao, S., Jia, L., Sun, G., Wang, R., and Liang, Y.
\newblock Tileflow: A framework for modeling fusion dataflow via tree-based analysis.
\newblock In \emph{Proceedings of the 56th Annual IEEE/ACM International Symposium on Microarchitecture}, pp.\  1271--1288, 2023.

\bibitem[Zhou \& Yang(2022)Zhou and Yang]{zhou2022exploring}
Zhou, Y. and Yang, K.
\newblock Exploring tensorrt to improve real-time inference for deep learning.
\newblock In \emph{2022 IEEE 24th Int Conf on High Performance Computing \& Communications; 8th Int Conf on Data Science \& Systems; 20th Int Conf on Smart City; 8th Int Conf on Dependability in Sensor, Cloud \& Big Data Systems \& Application (HPCC/DSS/SmartCity/DependSys)}, pp.\  2011--2018. IEEE, 2022.

\end{thebibliography}
